\pdfoutput=1

\documentclass[8.5pt,twoside,twocolumn]{article}
\usepackage[super,sort&compress,comma]{natbib} 
\usepackage{mhchem}
\usepackage{times,mathptmx}
\usepackage{natbib}

\usepackage{sectsty}
\usepackage{balance} 

\usepackage[pdftex]{graphicx}  %
\usepackage{epstopdf} %
\usepackage{pslatex} 
\usepackage{lastpage}
\usepackage[format=plain,justification=raggedright,singlelinecheck=false,font=small,labelfont=bf,labelsep=space]{caption} 
\usepackage{fancyhdr}
\usepackage{amssymb}

\begin{document}

\thispagestyle{plain}
\fancypagestyle{plain}{
\fancyhead[L]{\includegraphics[height=8pt]{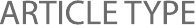}}
\fancyhead[C]{\hspace{-1cm}\includegraphics[height=20pt]{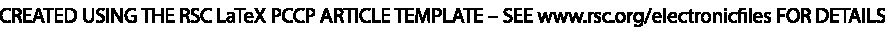}}
\fancyhead[R]{\includegraphics[height=10pt]{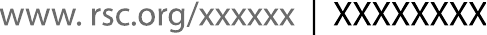}\vspace{-0.2cm}}
\renewcommand{\headrulewidth}{1pt}}
\renewcommand{\thefootnote}{\fnsymbol{footnote}}
\renewcommand\footnoterule{\vspace*{1pt}%
\hrule width 3.4in height 0.4pt \vspace*{5pt}} 
\setcounter{secnumdepth}{5}

\makeatletter 
\def\subsubsection{\@startsection{subsubsection}{3}{10pt}{-1.25ex plus -1ex minus -.1ex}{0ex plus 0ex}{\normalsize\bf}} 
\def\paragraph{\@startsection{paragraph}{4}{10pt}{-1.25ex plus -1ex minus -.1ex}{0ex plus 0ex}{\normalsize\textit}} 
\renewcommand\@biblabel[1]{#1}            
\renewcommand\@makefntext[1]%
{\noindent\makebox[0pt][r]{\@thefnmark\,}#1}
\makeatother 
\renewcommand{\figurename}{\small{Fig.}~}
\sectionfont{\large}
\subsectionfont{\normalsize} 

\fancyfoot{}
\fancyfoot[LO,RE]{\vspace{-7pt}\includegraphics[height=9pt]{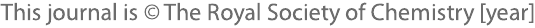}}
\fancyfoot[CO]{\vspace{-7.2pt}\hspace{12.2cm}\includegraphics{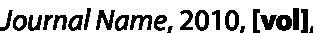}}
\fancyfoot[CE]{\vspace{-7.5pt}\hspace{-13.5cm}\includegraphics{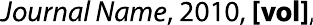}}
\fancyfoot[RO]{\footnotesize{\sffamily{1--\pageref{LastPage} ~\textbar  \hspace{2pt}\thepage}}}
\fancyfoot[LE]{\footnotesize{\sffamily{\thepage~\textbar\hspace{3.45cm} 1--\pageref{LastPage}}}}
\fancyhead{}
\renewcommand{\headrulewidth}{1pt} 
\renewcommand{\footrulewidth}{1pt}
\setlength{\arrayrulewidth}{1pt}
\setlength{\columnsep}{6.5mm}
\setlength\bibsep{1pt}

\twocolumn[
  \begin{@twocolumnfalse}
\noindent\LARGE{\textbf{Energies and pressures in viruses: contribution of nonspecific electrostatic interactions$^\dag$}}
\vspace{0.6cm}

\noindent\large{\textbf{Antonio \v{S}iber,$^{\ast}$\textit{$^{a,c}$}, An\v ze Lo\v sdorfer Bo\v zi\v c{$^{b}$} and
Rudolf Podgornik\textit{$^{b,c,d}$}}}\vspace{0.5cm}

\noindent\textit{\small{\textbf{Received Xth XXXXXXXXXX 20XX, Accepted Xth XXXXXXXXX 20XX\newline
First published on the web Xth XXXXXXXXXX 200X}}}

\noindent \textbf{\small{DOI: 10.1039/b000000x}}
\vspace{0.6cm}

\noindent \normalsize{We summarize some aspects of electrostatic interactions in the context of viruses. A simplified but, within well defined limitations, reliable approach is used to derive expressions for electrostatic energies and the corresponding osmotic pressures in single-stranded RNA viruses and double-stranded DNA bacteriophages. The two types of viruses differ crucially in the spatial distribution of their genome charge which leads to essential differences in their free energies, depending on the capsid size and total charge in a quite different fashion. Differences in the free energies are trailed by the corresponding characteristics and variations in the osmotic pressure between the inside of the virus and the external bathing solution. 
}
\vspace{0.5cm}
 \end{@twocolumnfalse}
  ]

\section{Introduction}
\label{sec:into}

\footnotetext{\dag~Electronic Supplementary Information (ESI) available: [details of any supplementary information available should be included here]. See DOI: 10.1039/b000000x/}

\footnotetext{\textit{$^{a}$~Institute of Physics, Zagreb, Croatia and Department of Physics, Faculty of mathematics and physics, University of Ljubljana, SI-1000 Ljubljana, Slovenia.; E-mail: asiber@ifs.hr}}
\footnotetext{\textit{$^{b}$~Department of Theoretical Physics, Jo\v{z}ef Stefan Institute, SI-1000 Ljubljana, Slovenia. }}
\footnotetext{\textit{$^{c}$~Department of Physics, Faculty of mathematics and physics, University of Ljubljana, SI-1000 Ljubljana, Slovenia. }}
\footnotetext{\textit{$^{d}$~Department of Physics, University of Massachusetts, Amherst, MA  01003, USA. }}

Viruses are abundant and ubiquitous \cite{ubiquitous} and it is possible that there are no forms of life immune to the effect of viruses \cite{Holmes}, which may be advantageous in the fight against 
disease \cite{bacteriotherapy1,bacteriotherapy2,nancy}. It appears that there are even viruses that initiate their "lifecycle" exclusively in combination with some {\em other} viruses, often "stealing" the protein material 
of those viruses and diverting the cellular processes they initiated to their own advantage \cite{Baker_review,Rossmann} - the viruses are 
thus parasites even of their own kin. Although we know the exact nature (the shape and the genome) of only about a hundred \cite{viper} viruses, it seems that they are almost as diverse as life itself, so that they represent a type of index to the library of life forms that they parasitize upon. The viruses are indeed only abbreviated, indexed, crippled representation of life and they can hardly be classified as life. They are most often viewed by physicists (or "physical virologists" \cite{Pollard,physical_virology_book,bruinsma_nature_review,zlotnick_PNAS} \footnote{The first Gordon conference on "physical virology" was held in 2009
in Galveston, Texas.}) as hetero-macromolecular complexes, i.e. complexes of viral proteins and the genome molecule (DNA or RNA) that are reasonably stable in extra-cellular conditions and that initiate a complicated sequence of molecular interactions and transformations once they enter a suitable cell \cite{Wimmer}. According to such a view, a virus must in its structure somehow "encode" the crucial steps of its replication process. For example, the proteins that make its protective shell (virus capsid) must have such geometric and chemical characteristics as to activate the appropriate receptors on the cell membrane so that they can attach to and penetrate its interior \cite{principles_Cann,evilevitch}. The reverse of this process when mature viruses are released from the infected cell implies membrane adsorbtion and budding  \cite{benshaul,desernobudding} to a large extent promoted by electrostatic interactions \cite{wong}. The virus needs to be sufficiently stable in the extracellular conditions, yet sufficiently unstable once it enters the cell, so that it can disassemble and deliver its genome molecule to the cellular replication machinery \cite{Arkhipov}. Once it fulfills walking this tight rope of incipient instability, the manufacturing of virus components in the cell proceeds, leading eventually to new viruses \cite{roya_nucleation}. It is certainly of interest to elucidate the nature of interactions in viruses that enable it to function "between a rock and a hard place", equilibrating on the border of stability, not just from a fundamental scientific point of view but also technologically \cite{steinmetz1,steinmetz2}.

In this review, we shall concentrate on the description of virus structure in terms of the electrostatic interactions, i.e. we 
shall be interested in the corresponding {\em energies} of single viruses (e.g. the energy required to assemble a virus from its constituents) and osmotic
{\em pressures} acting in a virus, though electrostatic interactions are just as important  for understanding the interactions {\sl between} the virues \cite{Millman}.
In fact, the relevance of strong electrostatic interactions for the stability of tobacco mosaic virus (TMV) gels was invoked already in the seminal work of Bernal and Fankuchen \cite{Bernal}\footnote{They refer to electrostatic interactions as being "probably due to the ionic atmospheres surrounding [viruses]".}, while the Poisson-Botzmann theory of electrostatic interactions was applied to the case of viruses \cite{Oster} soon after its publication \cite{Verwey} and even before it was applied to lipid membranes \cite{Parsegian-thesis}.

We shall examine the relation of the formation energy and osmotic pressure of a virus  to its structure. Our emphasis is on the electrostatic interaction, first because it contributes a significant part to overall energetics of the viruses \cite{Konecny,Shields,parsegian_bacteriophage,Jiang} but also because the electrostatic part of the free energy is the part that can be calculated with a fair amount of precision \cite{Evans}. Subtler properties of viruses, such as the nature and extent of the ordering of packed DNA/RNA molecule \cite{Marenduzzo,Marenduzzo2} may depend on other physical effects, such as genome molecule entropy, yet an important contribution to the energy of protein-genome packaging is of electrostatic nature \cite{Marenduzzo3}.

The aim of this work is to show the usefulness of elementary physical concepts describing electrostatic interactions as they apply to viruses \cite{phages}. 
Although we will explain some of the intricacies involved in the modeling of viruses, our emphasis will be on the essential physics involved. 
Although the simple (and necessarily approximate) expressions that we shall expose and employ in this review can be derived from previous, more 
elaborate publications, we intend to use mostly scaling-based reasoning in their derivation and application in the context of viruses.

\begin{figure}[t!]
\centering
  \includegraphics[height=10cm]{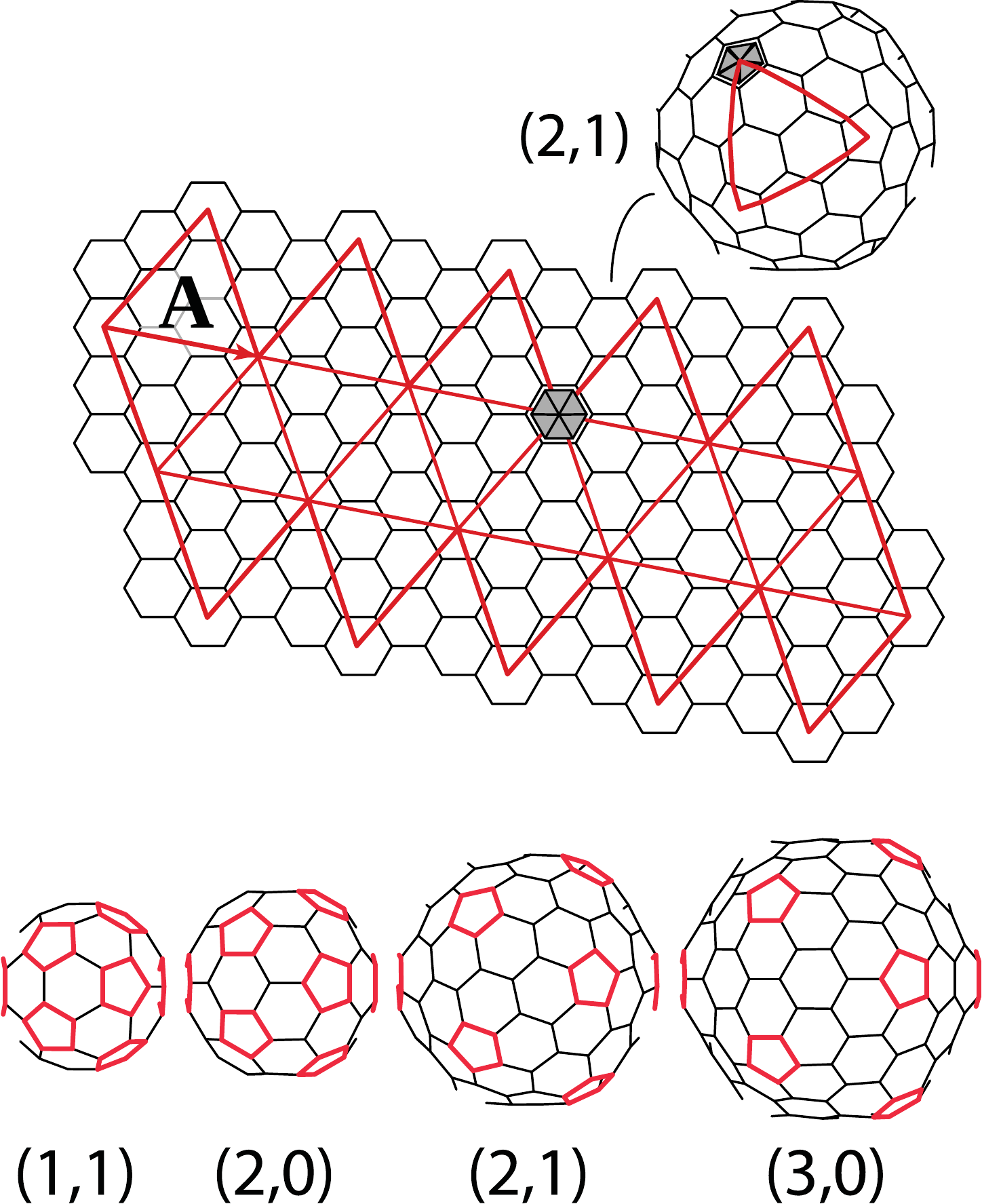}
  \caption{The geometry behind the Caspar-Klug (CK) classification of viruses. Icosahedral viruses that obey the CK principle can be 
"cut out" of the lattice of protein hexamers, as shown in the figure. Upon folding of the cut-out piece, twelve of protein hexamers are 
transformed in pentamers. The CK viruses are described with two integers, $h =2$ and $k=1$) in the case shown, which parametrize the vector 
${\bf A}$. The $T$-number of the capsid is related to $h$ and $k$ as $T=h^2 + hk + k^2$, and the number of protein subunits is $60 T$; 
see Refs. \cite{Baker_review,siber_arxiv} for details. The bottom row of images displays the CK structures with $T=3,4,7$, and 9 (from left to right).}
  \label{fig:figCK}
\end{figure}

Electrostatic interactions in the context of viruses are complex and bring together various theoretical approaches extending from the theory of 
polyelectrolytes \cite{Dobrynin,Andelman_Netz} and then all the way to the theory of highly charged Coulomb fluids \cite{naji-rev,hoda-rev}, sometimes exploiting elaborate theoretical concepts and formally demanding approaches. In this review we shall thus try to retain only most easily understood and applicable concepts  \cite{Andelman_pap}, yet sufficiently reliable when it comes to description of biological system in general and viruses in particular \cite{Rudi_book,Poon_book}. This approach necessarily excludes the details of some of the more arcane aspects of the theory of electrostatic interactions in aqueous solutions (non-linear salt screening, effects of polyvalent counterion correlations, overcharging by the mobile charge and similar \cite{reviewES}). We do explain the essence of these aspects and their possible relevance in context of viruses in Section \ref{sec:strong}.

We shall emphasize the {\em non-specific} aspects of the contributions electrostatic interactions and will thus avoid a detailed exposition of the 
ion-specific effects \cite{Ben-Yaakov1,Ben-Yaakov2}. This means that we do not deal with chemical specificity of different ions that may drastically modify electrostatic interactions. These effects may be of importance for initiation of assembly, or for speeding up of the assembly, but they are not of primary interest to this review. Our point of view concerns the more robust aspects of virus energetics that can be understood 
in a sufficiently generic (and simple) physical framework.

\section{A simple description of a virus}
\label{sec:simple_virus}

The emphasis of this review is not on the symmetry and shapes of viruses - there are already many good reviews on this subject (see e.g. Ref. \cite{Baker_review}). For our purposes it is enough to state that all viruses are made of two essential parts: protein coating or a {\em capsid} and viral genome (of DNA or RNA type) situated in the capsid interior. There are also viruses that in addition to these two essential components need an additional "wrapper", i.e. a piece of cellular membrane, in order to function properly and fuse with the cellular membrane surface \cite{fusion}. These viruses are referred to as {\em enveloped} (in contrast to {\em non-enveloped} viruses which do not have a membrane coating). Because of severe restrictions on the length of their genome encoding the viral shell proteins, the virus capsids are made of {\em many copies} of one or at most a few types of proteins which are arranged in a highly symmetrical manner as first proposed in the seminal work by Crick and Watson \cite{Crick}.

Nearly spherical viruses, also called {\em icosahedral viruses}, show mostly but not always \cite{roya_origin} icosahedral order and the proteins that make them can be arranged in the clusters of five (pentamers) or six (hexamers; see Fig. \ref{fig:figCK} for details and definition of Caspar-Klug classification) \footnote[1]{There are also viruses with icosahedral order, yet containing only pentamers. Such is the polio virus \cite{Baker_review}}. This arrangement may be only conceptual, but may also have a physical meaning that the interactions in clusters (capsomeres) are somewhat stronger than the interactions between the clusters \cite{ceres,Baker_review}. Crick and Watson surmise that nearly all viruses can be classified either as nearly spherical, i.e. of icosahedral symmetry, or elongated of helical 
symmetry \cite{Crick}. 

Icosahedral viruses tend to look more polyhedral when larger \cite{Baker_review,LMN,siber_inpressure}. 
There are also non-icosahedral viruses that do not fit in CK classification. Here are some prominent examples: Capsids of some bacteriophages (viruses that infect only bacteria) are "elongated" (prolate) icosahedra \cite{Luque}, i.e. the icosahedral sides around the equator are not equilateral, but isosceles triangles \cite{prolate_icosahedron}. Capsids of some plant viruses (e.g. tobacco mosaic virus) are (open and hollow) cylinders and their genome molecule is situated in the empty cylindrical space formed by proteins. HIV virus is also non-icosahedral, but is not an elongated icosahedron \cite{Roya_HIV}. Its capsid typically looks conical, being elongated and narrower on one side \cite{HIV-JMB}. Furthermore, even when the viruses are spherical, it is sometimes difficult to classify them according to CK scheme and the typical 
pentamer-hexamer ordering is not evident\cite{Lorman1,Lorman2}. Some viruses are multi-layered, i.e. they consist of several protein capsids each of which may be 
built from different protein \cite{bluetongue_multi}. Each of these capsid layers may individually conform to the CK principle \cite{aquareovirus}. 
An alternative to CK classification has recently been proposed that apparently contains the CK shapes as the subset of all 
possible shapes, including those that do not show a clear pentamer-hexamer pattern \cite{Lorman1,Lorman2}.

\begin{figure*}[t]
\centering
  \includegraphics[width=18cm]{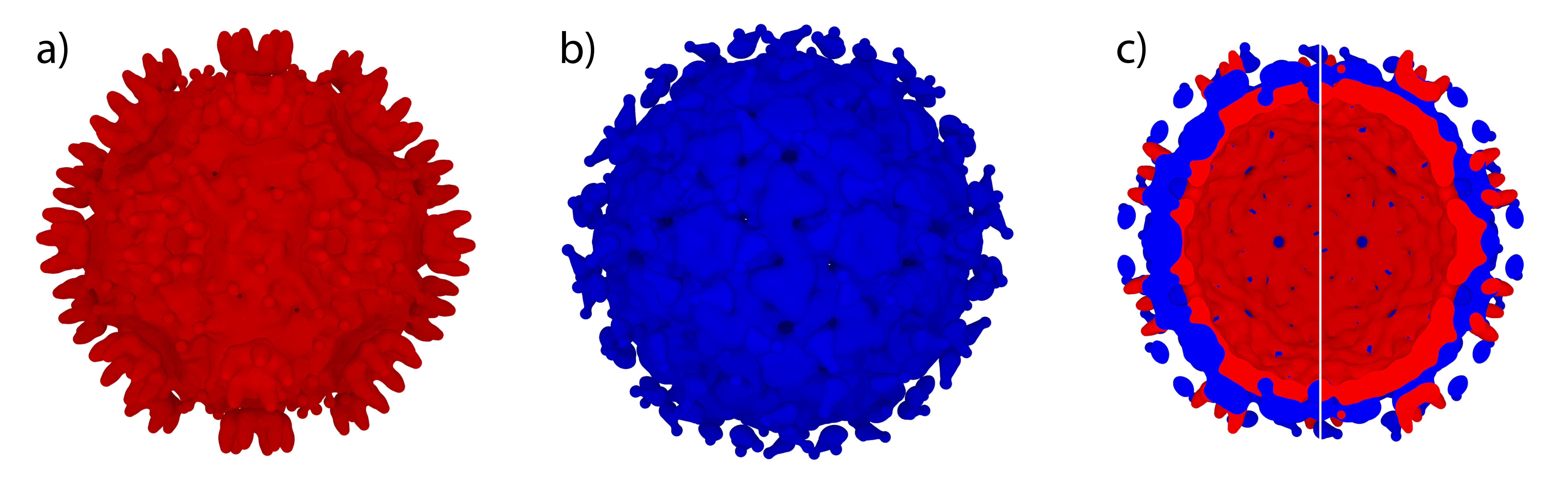}
  \caption{The calculated representation of the charge distribution on the capsid of cowpea chlorotic mottle (ssRNA) virus as explained in the text, based on the 1CWP entry in the 
RCSB Protein Data Bank: a) The isosurface of positive charge (red). b) The isosurface 
of negative charge (blue). c) The combined isosurfaces of positive 
and negative charges shown in the capsid cut in half so that its interior is seen. On the lef-hand side of image in panel c) (the left of the white vertical line), the (cut) isosurface of negative 
charge (blue) is translated infinitesimally closer to the viewer, while it is the opposite on the right-hand side of the image.}
  \label{fig:charges}
\end{figure*}

There is a certain universality in the size of capsid proteins. By analyzing more than 80 different viruses (with $T$ numbers from 1 to 25), 
we have found that the area of a protein in a capsid is fairly conserved and amounts to $\sim$ 25 nm$^2$ \cite{sib_anze_rudi_unpub1}. The thickness of the protein, i.e. the thickness of the virus capsid in question varies more, but is typically in the interval $\sim$ 2 - 5 nm. The "typical" virus protein can thus be imagined as a disk / cylinder of radius $\sim$ 3 nm and thickness $\sim 3$ nm. In some viruses these "disks" have positively charged protein "tails" that protrude in the capsid interior and whose role is to bind to a a negatively charged genome molecule (typically ssRNA; see Sec. \ref{sec:energies_ssRNA} and Ref. \cite{muthukumar}). There is some universality in the distribution of charges along and within the virus capsid, Fig. \ref{fig:charges}  
\footnote{The iso-surfaces obtained here were calculated by first assigning one elementary positive charge to charge residue of {\em each} lysine and arginine amino acid and protein N-termini, one negative charge to charge residue of each aspartic acid and glutamic acid and protein C-termini, and 0.1 positive charges to histidine charge residues (this corresponds to the histidine fractional charge at neutral p$H$). We thus obtained the 3D coordinates of charges in the capsid, yet this discrete data does not give a visually clear insight in the three-dimensional nature of the charge distribution. To obtain an informative visual representation, we used this data to construct spatial scalar fields of positive and negative 
charge distributions (separately), by assigning to each charge residue a scalar density, $\tau (r) = |q/e_0| \left \{ 1 - [\rm{min}(r,W) / W]^2 \right \}^2$, where $q$ is the charge of the residue, $e_0$ is the elementary charge, $r$ is the distance from the charge residue, and $W$ is the parameter specifying the extent of the density. The total density field was constructed by summing the scalar densities of all the charges (separately for positive and negative distributions). From the thus obtained space field, an isosurface showing the value of the field at $t$ is plotted. The objects obtained in this way are known in the computer graphics community as metaballs or blobs \cite{Blinn}. In Figs. \ref{fig:charges} and \ref{fig:fig_virus_tails}, $W=1.34$ nm and $t=0.85$. Amino acids in the protein interior, i.e. in the capsid wall may not carry charge, as we have assumed, but due to the nature of graphical representation we have chosen, panels a) and b) of Fig. \ref{fig:charges} (but not panel c) would ad equately represent the (surface) charge distributions also in the case when only amino acids sufficiently close to the capsid surface were charged.}. While Caspar-Klug dipoles, corresponding to a bimodal distribution of positive charges on the hypotopal side and negative charges on the epitopal side of the capsid can be observed \cite{kegel_schoot}, and we recognize them also for the CCMV capsid in Fig. \ref{fig:charges}, this is certainly not a rule and other, e.g. monomodal, distributions can be observed just as well. The in-plane angular distribution of charges along the capsid thickness also shows complicated variations within the constraints of the icosahedral symmetry group \cite{sib_anze_rudi_unpub1}, see Fig. \ref{fig:charges}. Last but not least, the magnitude of the charges on the surface of the capsomeres is regulated by the dissociation equilibrium while for the buried charges it would have to be estimated from quantum chemical calculations \cite{Ching,Qprot}.

Thus far, we have talked about the order of virus proteins and we have said nothing about the distribution and ordering of the virus DNA or RNA 
molecule inside the capsid. This will be discussed in the following sections. The virus genome molecule codes for the proteins of the capsid, but 
also for other proteins needed in the process of virus replication, depending on a virus in question. ssRNA viruses need to code for protein that replicates the virus ssRNA and some viruses also encode the regulatory proteins that are required for correct assembly (scaffolding proteins, see e.g. Ref. \cite{scaffold1}) and the proteins required for release of viruses from the infected cell \cite{principles_Cann}. The amount of information that is required constrains the length of the genome molecule from below.

\section{Self-assembling viruses and their energies}
\label{sec:self_assembly}

Many viruses can self-assemble \cite{Bancroft2} though the details of the assembly pathways are seldom well understood \cite{Chandler}. This means that the "ingredients" for a virus, individual proteins and its genome molecule, can spontaneously form (assemble in) closed, functional viruses, even outside the cellular environment, in conditions of appropriate pH factor and salt concentration. This is typical for viruses that contain ssRNA molecule and it was first demonstrated in tobacco mosaic virus by Fraenkel-Conrat and Williams \cite{fraenkelconrat}. They were able to produce infectious virus particles by simply mixing two solutions, one of them containing only virus proteins, and the other virus RNA molecule.

The process of self-assembly can proceed spontaneously only if the free energy ($F$) of the assembled virus is lower than the free energy of the disassembled state. This means that, some viruses at least, can be viewed as thermodynamically optimal structures, i.e. they represent minima of the free energy \cite{Bruinsma_thermoPRL}.

Upon assembly of proteins and RNA in a virus, their entropy ($S$) decreases, so the process can not proceed spontaneously unless there is a gain in the internal energy ($U$) of the system, i.e. there is a favorable "binding" energy of ingredients once they form a virus. Even then, the process of assembly can proceed only when the concentration of the ingredients is sufficiently large, i.e. above a critical concentration required for assembly \cite{siber_majdandzic}. Below that concentration, the entropic contribution to the free energy dominates and although the binding energy (enthalpy) of the ingredients in a capsid is favorable, the net gain in free energy is not, and the proteins and RNA molecules remain in a disassembled state.

Not all viruses can self-assemble. Some types of viruses, even when all the ingredients are available, can form only in the cellular environment, i.e. they require some of the cellular mechanisms for the assembly. The typical requirement is ATP energy which suggests that those viruses do not correspond to the simple free energy minima, but are rather examples of free energy driven structures at some elevated position/plateau in the free energy landscape. The use of ATP energy for assembly is typical for bacteriophage viruses as we shall see in the following.

\subsection{Energies and assembly of empty viruses}

The viruses are kept together by the same interactions as those governing the "living matter" \cite{longrange,Leckband,nancy}. The interactions are thus many different guises of electromagnetic force, sometimes as "direct" interactions between entities (e.g. electrostatic, van der Waals (vdW), steric repulsion 
interaction,...)\cite{Daune} and sometimes as "indirect" or effective interactions which manifest as forces but arise only in presence of the bathing 
medium at finite temperature, $T$ (e.g. hydrophobic or hydration interactions) \cite{Israel,Ptitsyn}. The direct interactions are also modified by the presence and 
nature of the bathing medium, yet they survive outside it.

A very important characteristic of virus genome molecules is the {\em negative charge} they carry, due to dissociation of phosphate groups on 
RNA and DNA bases \cite{Bloomfield}. A compactification of a highly negatively charged genome in a capsid interior requires energy. That is why 
the proteins in contact with the DNA or RNA are often {\em positively charged}. This may not be the case for the protein as a whole, but it 
is quite often the case for parts of the surface of proteins in direct contact with the DNA or RNA \cite{karlin_charge,muthukumar} (see Fig. \ref{fig:charges}; note that the 
capsid interior is mostly red, i.e. positively charged). These parts of proteins are often spatially extended and they in some cases form highly positive "tails", see next section. The complementary charges on the viral proteins and genome increase the electrostatic interaction of the complex and decrease its energy, enabling an easier assembly. That is the reason the viral genome codes for such proteins, but there are viruses (e.g. polyomavirus) which utilize {\em cellular}, very positively charged proteins (histones) to associate them with their genome so as to reduce the electrostatic energy and allow for an easier packing \cite{principles_Cann}. All this indicates the importance of electrostatic interactions for virus assembly.

The charge on the proteins (and on the DNA/RNA) depends on the pH factor of the 
solution. Modifications of pH factor, and the corresponding dissociation/saturation equilibrium of ionic bonds in proteins leads to variation of their 
charge \cite{Oosawa_book,Ptitsyn} and a change in the electrostatic energy of the nucleoprotein complexes, as will be seen in the following. Modification of the 
pH factor may occur in the cellular environment in endosomes/vesicles i.e. pieces of cellular membrane that the viruses carry along with 
them as they penetrate the cell. Some viruses assemble in specific subcellular compartments (called "virus factories" \cite{virus_factories} 
or inclusion bodies\cite{principles_Cann}) which may also have an adequately altered, local pH value.

\subsubsection{A simple model of a virus capsid and the corresponding energy.~~}

The simplest representation of the protein charge distribution would be a positive charge homogeneously distributed on a protein. A homogeneous distribution of positive charge 
will tend to keep proteins apart, yet it has been experimentally observed that proteins of some viruses assemble in (empty) capsids under appropriate 
conditions. Such is a virus of hepatitis B that has been experimentally studied in considerable detail by A. Zlotnick and coworkers \cite{ceres}. 
Since the viral proteins do assemble in empty capsids, there must be an attractive contribution to their interaction. The source of this 
contribution is a combination of hydrophobic and van der Waals interactions\cite{Israel}, with hydrophobic component playing a dominant role. 
This conclusion can be read out from the experiments performed by Ceres and Zlotnick \cite{ceres}. They observed that the strength of protein-protein 
interactions in capsids {\em increases} with temperature and this suggests an important entropic contribution to the 
interaction, hydrophobic interaction being the obvious candidate \cite{kegel_schoot}.

To understand the cohesive energy of a virus capsid, we first need to evaluate a seemingly simple problem: obtain the electrostatic self-energy of 
a uniformly (positively) charged (with surface charge density $\sigma$), permeable, infinitely thin sphere of radius $R$ - this is the zeroth level 
description of a capsid.  This problem can be solved on the mean-field level by treating the ions as ideal gas, which adjust to the external potential and contribute to it via their charge density. This is the Poisson-Boltzmann (PB) approach which yields nonlinear differential equation for the electrostatic potential, $\phi$ \cite{Daune,Andelman_pap,Dan_David}. It can be derived by minimizing the appropriate free energy \cite{Safran} which has the form
\begin{equation}
F_{PB}[\phi(r), \nabla\phi(r), c^{i}(r)] =\int f_{PB}(\phi(r), \nabla\phi(r), c^{i}(r)) d^3r , 
\end{equation}
where the free energy density is given by
\begin{eqnarray}
& & f_{PB}(\phi(r), \nabla\phi(r), c^{i}(r)) = - {\textstyle\frac{1}{2}} \epsilon_0 \epsilon \nabla\phi(r)^2 + \sum_{i= \pm} e_{i} c^{i}(r) \phi(r) + \nonumber \\
&+& \sum_{i= \pm} k_B T \left [  c^{i}(r) \ln c^{i}(r) - c^{i}(r) - \left( 
c_0^{i} \ln c_{0}^{i} - c_{0}^{i} \right) \right ] + \nonumber\\
& &  + ~e_0 \rho_p(r) \phi(r) . 
\label{eq:fpb}
\end{eqnarray}
Here $e_0 \rho_p(r)$ is the charge density of the capsid, $T$ is temperature, $k_B$ is the Boltzmann constant, 
$c^{i}$ are the concentrations of (monovalent) salt ions, with $c_{0}^{i}$ their bulk concentrations, $\epsilon \epsilon_0$ is the permittivity of water, and $e_0$ is the electron charge.
Minimizing the above free energy w.r.t. $\phi(r), \nabla\phi(r)$ as well as $c^{i}(r)$, leads to the PB equation of the form
\begin{equation}
-\epsilon \epsilon_0 \nabla^2\phi(r) = \sum_{i= \pm} e_i c_0^{i} e^{- \beta e_i \phi(r)} + e_0 \rho_p(r),
\label{PBequ}
\end{equation}
where  $\beta^{-1} = k_B T$ and $e_i$ is the charge of the ions, i.e. $e_i = \pm e_0$.

When the electrostatic potentials in the solution are small, $e_0 \beta \phi \ll 1$ and we are dealing with a symmetric system, e.g. 1-1 electrolyte, that has $c_0^{+} = c_0^{-} = c_0$, the PB equation can be linearized yielding the Debye-H\"{u}ckel (DH) equation for the potential \cite{debyehuckel}, of the form 
\begin{equation}
- \nabla^2\phi(r) = \frac{\beta}{\epsilon \epsilon_0} \left( \sum_{i= \pm} e_i^2 c_0^{i}\right) \phi(r) + \frac{e_0 \rho_p(r)}{\epsilon \epsilon_0} + \dots,
\label{DHequ}
\end{equation}
where we took into account that the salt is assumed to be uni-univalent and thus $\sum_{i= \pm} e_i c_0^{i} = 0$. At this point one standardly introduces the inverse (Debye-H\"{u}ckel) screening length $\kappa^{-1}$,  with $\kappa^2 =  {\beta} \left( \sum_{i= \pm} e_i^2 c_0^{i}\right)/\epsilon \epsilon_0$ \cite{Verwey,Derajguin}.

The linearity of DH equation renders it amenable to several ways of solving, including the Green function method \cite{anze_sib_rudi}. But the simplest way to think of the DH approximation is in terms of the renormalization of electrostatic interactions (in vacuo) by salt ions. The effective interaction between the charges $Q_1$ and $Q_2$, separated  by ${\bf r}_1 - {\bf r}_2$ in the solution of monovalent ions (with concentration $c_0$) of (relative) dielectric constant $\varepsilon$ is given by the DH potential of the screened exponential form $U({\bf r}_1 - {\bf r}_2)$.
The easiest way to obtain the self energy of a uniformly charged sphere in the DH approximation is to sum the pair DH interactions over the sphere surface,
\begin{eqnarray}
F_{DH} &=& \frac {\sigma^2}{2} \int dS_1 \int dS_2 ~U({\bf r}_1 - {\bf r}_2)  = \nonumber\\
&=& \frac{1}{2} \frac{\sigma^2}{4 \pi \varepsilon_0 \varepsilon}  \int dS_1 \int dS_2 
\frac{\exp \left( -\kappa |{\bf r}_1 - {\bf r}_2| \right)}{|{\bf r}_1 - {\bf r}_2|},
\label{eq:DH_gen}
\end{eqnarray}
where $dS_1$ and $dS_2$ are infinitesimal elements of the sphere surface centered around vectors ${\bf r}_1$ and ${\bf r}_2$, and 
factor of $1/2$ accounts for double counting of the pair interactions. Since the radii of viruses are typically of the order of 20 nm, and the DH screening length in the physiological conditions ($c_0 \sim $ 150 mM) is $\kappa^{-1} \sim$ 1 nm, the limit of $\kappa R \gg 1$ is often implied. The range of integration is effectively cut on the scale of $\kappa^{-1}$. In case of interactions on a sphere, this defines a spherical cap. But, when $\kappa R \gg 1$, this spherical cap transforms in a disk, and this renders the two integrations independent:
\begin{equation}
\lim_{\kappa R \gg 1} F_{DH} = \frac{\sigma^2}{4 \pi \varepsilon_0 \varepsilon} \frac{1}{2} \int dS_1 \int_{0}^{\infty} dr_2 \int_{0}^{2\pi} d \phi_2
\exp \left( -\kappa r_2 \right).
\end{equation}
Note that the integration over $r_2$ can be extended to infinity, since $\kappa$ in the exponential function acts as a cutoff parameter \cite{anze_sib_rudi}. The self-energy of the capsid is thus 
\cite{kegel_schoot} 
\begin{equation}
\lim_{\kappa R \gg 1} F_{DH} = \frac{\pi \sigma^2 R^2}{\varepsilon_0 \varepsilon \kappa}.
\label{eq:DH_empty}
\end{equation}
We have denoted the self-energy using letter $F$ to indicate that it corresponds to free energy, containing also the entropy of salt ions. One should note the meaning of this quantity: it is the energy required to bring {\em infinitesimal} charges from infinite separations 
in the solution to the capsid. When the capsid consists of many weakly charged proteins, one may think of this quantity as being approximately the electrostatic contribution to the {\em assembly free energy}. To estimate this quantity, we need an information on the surface charge density of the virus proteins. This can be estimated from their amino acid content, and typically $\sigma \sim 1$ $e_0$/nm$^2$ \cite{kegel_schoot}. Taking now $R \sim$ 20 nm and $c_0 = 100$ mM, we obtain $F_{DH} \sim 10^4$ $k_B T$. 

The hydrophobic energy can be estimated from the area of proteins engaged in protein-protein contacts, i.e. from the volume slice around the proteins one water molecules thick. The capsid of 20 nm typically have $T=3$ triangulation number, i.e. they consist of 180 protein subunits. The length of protein contacts in such a capsid is about 3000 nm. This gives the total area exposed to the protein contacts of 
about 6000 nm$^2$ if we assume the capsid thickness of 2 nm. The energy of attractive protein interactions (hydrophobic {\em and} van der Waals) per unit area of the exposed protein surface is typically of the order of 10 mJ / m$^2$ \cite{hydro_est}. Multiplying the estimated exposed area with this energy, the estimated attractive hydrophobic interaction is obtained as, $F_{HP} \sim$ $10^4$ $k_B T$. This is of the same order of magnitude as the electrostatic repulsion and one is led to conclude that the interactions tending to dissolve a capsid and those keeping it together are in a tight balance 
(Ceres and Zlotnick have measured the free energy of the hepatitis B capsid to be $-\sim$ 1.5 $10^3$ $k_B T$ i.e. about 5-6 $k_B T$ per interprotein contact\cite{ceres}). The reason for this is most likely the fact that viruses, in addition to being able to assemble, need also to disassemble and deliver their genome molecule to the cell. In simple viruses this may be triggered by variation of pH and ionic concentration in different cell regions which increase the electrostatic repulsion in the capsid. Sufficiently large changes in pH lead to variation (increase) of protein charge \cite{kegel_TMV}, i.e. $\sigma$, while reduction of ionic concentration leads to decrease of $\kappa$, and as Eq. (\ref{eq:DH_empty}) shows, both effect modify the electrostatic self-energy of the capsid. 
Ceres and Zlotnick \cite{ceres} have experimentally demonstrated that the free energy of hepatitis B capsids increases with concentration of mono-valent salt. This may be 
interpreted as the screening of repulsive protein-protein electrostatic interaction \cite{kegel_schoot,sib_rudi_empty}.
A sufficiently large modification of the electrostatic interaction may lead to enthalpic instability of the capsid and to its disassembly \cite{anze_sib_rudi}. 
Experimental studies of virus assembly \cite{ceres,adolph,lavelle} clearly show the importance of pH factor and the salinity of the solution for the assembly of complete capsids. For sufficiently large concentration of virus proteins, the assembly depends on the values of both of these parameters \cite{adolph,lavelle}.

\subsubsection{Refined models of virus capsids.~~}

The simplest approach to capsid electrostatics presented above may be improved in several respects. The easiest one is to 
examine the complete DH solution to the problem, i.e. in a whole range of $\kappa R$ values, valid in particular for lower 
ionic concentrations when $\kappa < R$. In fact one can derive a general formula\cite{sib_rudi_empty}
\begin{equation}
F_{DH} = \frac{2 \pi \sigma^2 R^2}{\varepsilon_0 \varepsilon \kappa \left[ 1 + \coth( \kappa R) \right]}
\end{equation}
that describes the capsid electrostatic free energy for any $\kappa$ within the range of validity of the DH approximation..

A next level of refinement is to abandon the assumption of smallness of electrostatic potential $\beta e_0 \phi \ll 1$ and to 
solve the PB equation, Eq. \ref{PBequ},  for the potential and obtain the capsid energy in this way. This is still not the 
"exact" solution to the problem, since the PB approach is an approximation of the mean-field genre and it neglects ionic 
correlations \cite{naji-rev}. Yet, the PB approach is more reliable for large surface charge densities and smaller ionic 
concentration, i.e. in cases where $\phi$ is not necessarily small. In the limit $\beta e_0 \phi \ll 1$ it of course reduces back to the DH limit. 
The detailed study and comparison of PB and DH approaches to capsid electrostatics has been performed in Ref. \cite{sib_rudi_empty}. 
The PB capsid energies are always smaller from the corresponding DH values, yet the functional dependence of $F$ on 
$\sigma$, $R$, and $c_0$ is similar in both approaches, the $R^2$ dependence in particular. The DH results are very reliable 
quantitatively when $R \sim 20$ nm, $\sigma <$ 0.8 $e_0$/nm$^2$ and $c_0 > 50$ mM. 

\begin{figure}[h]
\centering
  \includegraphics[height=7cm]{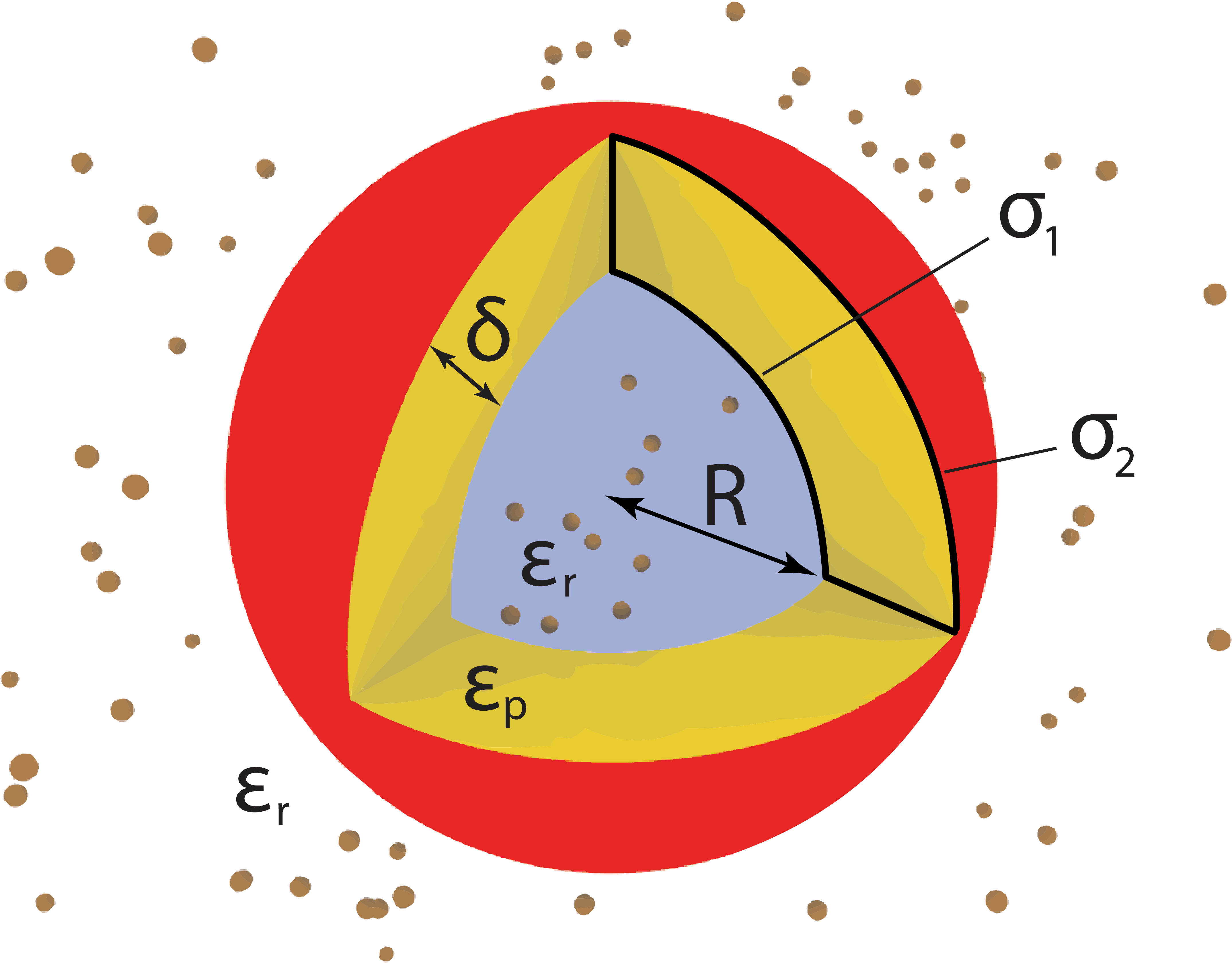}
  \caption{An illustration of an electrostatic model of a viral capsid with finite thickness, $\delta$. Salt ions are 
represented by small spheres. The interior of the capsid contains water and salt ions, but the salt ions are not present 
in the viral capsid shell.}
  \label{fig:fig_slab}
\end{figure}

The calculations that adopt a refined representation of the capsid with regards to its finite thickness, $\delta$, have also been performed. The distribution of charge across the thickness of the capsid is such that the positive charges are often concentrated on the capsid interior surface, while, typically negative charges are concentrated on the capsid exterior surface (see Fig. \ref{fig:charges}). There are also a few but functionally non-negligible \cite{garciamoreno1} net charges embedded in the interior of the capsid proteins \cite{Moreno,Ptitsyn}, while most of the charges there are partial charges due to electronic charge redistribution in chemical bond formation \cite{Qprot,chingbook}. To account for the finite thickness of the capsid, it was treated as a dielectric shell with relative permittivity $\varepsilon_p$, impermeable to ions, with interior and exterior surfaces which are uniformly charged \cite{sib_rudi_empty} with surface charge densities $\sigma_1$ and $\sigma_2$, respectively (see Fig. \ref{fig:fig_slab}). This model shall still prove useful, so we discuss it briefly in the following.

Intriguingly, for this model also, in the regime $\kappa R \gg 1$ the capsid free energy scales with the second power of capsid radius, both in the DH and 
PB cases \cite{sib_rudi_empty}. DH expressions for the electrostatic potential and free energy can be obtained analytically, yet they are sufficiently transparent only in certain limits. A limit 
of interest to us is $\kappa R \gg 1$, $\epsilon \gtrsim \epsilon_p$, $\delta \ll R$. In this case, 
\begin{equation}
F_{DH} (\sigma_1, \sigma_2, \delta) = 2 \pi R^2 \frac{\varepsilon_p \left( \sigma_1 + \sigma_2 \right)^2 + 
\varepsilon \left( \sigma_1^2 + \sigma_2^2 \right) \kappa \delta}{\varepsilon_0 \varepsilon \kappa 
\left( 2 \varepsilon_p + \varepsilon \kappa \delta \right)}.
\label{eq:F_diel_slab}
\end{equation}
For viruses in physiological conditions, we may further take $\varepsilon_p \ll \varepsilon$ (for proteins, 
$\varepsilon_p \sim$ 5 \cite{Ptitsyn}), and $\kappa \delta \sim 1$, which simplifies Eq. (\ref{eq:F_diel_slab}) to
\begin{equation}
F_{DH} = \frac{2 \pi (\sigma_1^2 + \sigma_2^2) R^2}{\varepsilon_0 \varepsilon \kappa}.
\end{equation}
By comparing this equation with Eq. {\ref{eq:DH_empty}) one concludes that the qualitative behavior of the capsid free energy is not 
importantly modified by an introduction of the finite capsid thickness and that the free energy is of the same order of magnitude as 
in the case of infinitely thin shell.

Angularly nonuniform distribution of the capsid charge (still positioned on a perfectly spherical, infinitely 
thin capsid), in accordance with its 
icosahedral symmetry, can be introduced in the electrostatic model. This complicates the treatment and introduces special function series, but 
the electrostatic free energy still retains the already familiar functional DH behavior in the $\kappa R \gg 1$ limit. 
In particular, Marzec and Day \cite{marzec_day} have obtained that
\begin{equation}
F = \frac{\pi \Sigma^2_Q R^2}{\varepsilon_0 \varepsilon \kappa},
\end{equation}
where $\Sigma^2_Q$ is the average square charge density, $\Sigma^2_Q \equiv (4 \pi)^{-1} \int d \Omega \Sigma^2_Q (\Omega)$, and 
$\Omega$ is the spatial angle.

A completely numerical PB approaches are also possible. These start with a determination of the spatial distribution of 
capsid charge based on the capsid amino acid content and the atomic coordinates determined from X-ray studies \cite{Konecny}. The 
PB equation is then solved on a three-dimensional grid using advanced numerical 
routines \cite{Konecny}. Such approaches are, however, less transparent concerning the scaling of energies with various parameters of 
the system, e.g. virus radius, its total charge, salt concentration, and similar.

\section{Energies of ssRNA viruses}
\label{sec:energies_ssRNA}

Viruses that contain the single-stranded RNA molecule (ssRNA) often self-assemble \cite{Bancroft}. A simplified description of the ssRNA molecule in these viruses characterizes it as a generic flexible polyelectrolyte \cite{deGennes_book,Oosawa_book}. The ssRNA flexibility is described by its effective {\em persistence length} i.e. the length on which the ssRNA refuses to bend. For flexible polyelectrolytes, this effective length is of the order of monomer (nucleotide) separation ($a \sim$ 0.5 nm) and they can thus be well described within the framework of the Edwards - de Gennes {\em flexible chain model} \cite{Edwards_book,degennes-rev}. It is the connectivity of the chain that gives an essential imprint to the behavior of monomers of the chain as opposed to free particles in solution \cite{Muthu-polyel}. The long range interactions between monomers, such as Coulomb interactions in the case of ssRNA, are modified in an essential way when coupled to the connectivity of the chain \cite{rudi_polyel}. The main difference between the flexible chain model describing the salient features of the ssRNA chain and the semiflexible chains, such as dsDNA, is that the elasticity of the chain is purely entropic \cite{Treloar} in the former while being enthalpic in the latter case \cite{Rubinstein,Swigon}. The connectivity of the chain introduces many important features also in the behavior of the chain in external fields as is the case in adsorption to charged surfaces \cite{rudi_poly_2} and the consequent bridging interactions present between two apposed charged surfaces \cite{rudi-PE}. 

It is an experimental fact \cite{pariacoto,nodavirus} (and any reasonable theory should account for it) that the ssRNA viruses contain the 
ssRNA molecule in a thin shell closely distanced from the interior capsid radius. In the 
flexible polyelectrolyte theory the thickness of this adsorbed ssRNA polyelectrolyte layer is of the order of $a$ \cite{Woodward}. 
The thickness of this ssRNA shell results in general from a relatively complicated interplay of all the interaction energies and chain entropy involved in the problem \cite{sib_rudi_ssRNA}: the {\em electrostatic contribution} to the interactions between negatively charged ssRNA and positively charged hypotope of the capsid and the {\em entropic contribution} of the constrained polyelectrolyte to the free energy \cite{Akesson,Turesson}. 

\subsection{Electrostatic interactions and energies of ssRNA viruses}

The total  free energy in this case is composed of the electrostatic interactions due to ions and charges on the polyelectrolyte, as well as the entropy of the polyelectrolyte chain  \cite{rudi_polyel,sib_rudi_ssRNA}.  The electrostatic part of the free energy, $F_{es}$, contains the PB functional, Eq. (\ref{eq:fpb}), augmented by the presence of the charges on the polyelectrolyte chain expressed in terms of the polyelectrolyte monomer concentration, $\rho(r)$ and reads \cite{borukhov}
\begin{eqnarray}
F_{es}[\phi(r), \nabla\phi(r), \rho(r)] &=& \int f_{es}(\phi(r), \nabla\phi(r), \rho(r)) d^3r ~- \nonumber\\
& & - ~\mu \left ( \int d^3 r \rho(r) - N \right ), 
\end{eqnarray}
where $\mu$ is the Lagrange multiplier enforcing the condition of fixed number of monomers, $N$, of the polyelectrolyte chain, with
\begin{equation}
f_{es}(\phi(r), \nabla\phi(r), \rho(r)) = f_{PB}(\phi(r), \nabla\phi(r))  - p e_0 \rho(r) \phi(r). 
\label{eq:fions}
\end{equation}
Here $f_{PB}(r)$ was defined in Eq. \ref{eq:fpb} and $pe_0$ is the charge per monomer, with $e_0$ is the electron charge and $0 < p < 1$. The part of the free energy due to the entropy of the flexible polyelectrolyte can be approximated in the so-called {\sl ground state dominance} \cite{deGennes_book,degennes-rev} as 
\begin{equation}
F_{ent}[ \rho(r), \nabla \rho(r)] = k_B T \frac{a^2}{6} \int d^3 r \frac{\left[ \nabla \rho(r) \right]^2}{\rho(r)}.
\end{equation}
Minimizing the sum of the electrostatic and entropic contributions then leads to a polyelectrolyte PB equation that can be solved numerically in the spherical geometry of the capsid \cite{sib_rudi_ssRNA}.  

Numerical solutions of the polyelectrolyte PB theory are complicated but they invariably point to the existence of an adsorption layer next to the internal positively charged wall of the capsid \cite{sib_rudi_ssRNA}. %
This leads to the conclusion that ssRNA should be non-uniformly distributed within the capsid, showing a relatively dense surface layer and a depleted core. The existence of the adsorbed layer along the periphery of the capsid then engenders the attractive polyelectrolyte bridging interactions of the type observed to act between planar charged surfaces \cite{rudi-PE,Abraham} but here act between different parts of the spherical hypotope and thus stabilize the protein shell.

\subsubsection{Scaling approach to the ssRNA packing inside a capsid.~~}

A scaling estimate of the electrostatic free energy of ssRNA packing inside a capsid can again be obtained in a suitably simplified framework and in appropriate range of parameters. It has been found that the optimal virus configuration in the physiological regime is such that the total charge on the encapsidated ssRNA is comparable to the charge on the capsid, being equal as the salt concentration decreases \cite{sib_rudi_empty,angelescu_bruinsma,muthukumar}. This is relatively easy to understand. When the charges on the capsid and on the ssRNA molecule are equal (but of the opposite signs), the ssRNA completely screens the protein charges, so that the salt ions almost need not to redistribute at all, especially when the ssRNA and the capsid can be brought in close contact. In all other cases, there is an effective, remaining charge, that the salt ions must screen by rearranging, increasing the total electrostatic energy of the system in this way. This 
simple argument is complicated by the fact that the ssRNA consists of {\em connected} charges - it is a polyelectrolyte molecule. 
This line of reasoning applies only to the total charge on the ssRNA and the capsid. When each nucleotide is assumed to carry a fixed charge, this also fixes the total length of the ssRNA molecule. There are studies, however, which predict that not all ssRNA bases carry an elementary charge \cite{muthukumar}. This depends on the dissociation equilibrium and charge regulation \cite{Ninham}, but it does not importantly influence simple energy estimates to be presented below. Similar arguments can be used also for encapsidated polyelectrolyte cargo different from ssRNA \cite{roya_cargo,Hagan_elec}.

The ssRNA configuration inside the capsid as obtained from the full polyelectrolyte PB theory  
that contain also the polyelectrolyte entropy can be approximated by two concentric spherical shells with opposite charge, $\sigma \equiv \sigma_2 \approx -\sigma_1$, separated by $\sim a$ (see Fig. \ref{fig:fig_ssRNA_density}). Thus, Eq. (\ref{eq:F_diel_slab}) 
should be of use to estimate the free energy of such a configuration. The important difference of the ssRNA virus with respect to the 
dielectric slab model of the capsid is that the space between the ssRNA and the proteins is permeable to salt and water. However, 
when $a \lesssim \kappa^{-1}$, the salt induced screening of ssRNA-capsid protein electrostatic interaction is incomplete, and the 
free energy of salt ion distribution in between the ssRNA and the capsid is small compared to the analogous contributions 
in the rest of the space. 

Thus, a simplest estimate of the electrostatic complexation free energy ($\Delta F_{C}$) of a ssRNA virus in the DH approximation 
can be obtained from Eq. (\ref{eq:F_diel_slab}), using $\varepsilon_p \approx \varepsilon$ and $\kappa \delta \sim 1$ 
(in physiological conditions) and subtracting the electrostatic self-energy of the capsid, Eq. \ref{eq:DH_empty}. 
This yields
\begin{equation}
\Delta F_C \approx F_{DH}(-\sigma, \sigma, a) - F_{DH}(0, \sigma, 0) \approx \frac{\pi \sigma^2 R^2}{\varepsilon_0 \varepsilon} \left( fa - \kappa^{-1}\right),
\label{eq:DH_ss}
\end{equation}
where $f$ is a numerical factor between 4/3 and 2 (it depends on $\kappa$ and $\delta$). Note here that the sign of this free energy difference depends on 
the competition between two length scales: $a$ and $\kappa^{-1}$. 
 
The above result is not very accurate in high salt concentrations, when $\kappa^{-1} < a$ and does not include the non-electrostatic 
self-repulsion contribution of the polyelectrolyte which can be shown to be smaller then the leading electrostatic contribution \cite{sib_rudi_empty}. 
Using the "typical" virus parameters and $a=0.5$ nm, we obtain $\Delta F_C \sim 0$, since $fa - \kappa^{-1} \sim 0$. A more detailed 
calculation \cite{sib_rudi_ssRNA} gives $\Delta F_C$ which is negative and about a quarter of the free energy of an empty 
capsid (at physiological conditions) $F_{DH}(0, \sigma, 0)$, consistent with lower bound on $f$. This signifies that 
spontaneous encapsidation of ssRNA is a delicate process that may even be suppressed in thermodynamical equilibrium, so that only {\em empty} capsids form. Indeed, formation of empty capsids and free ssRNA was theoreticaly predicted in the regime of high ionic concentrations, but also when the charge on the ssRNA is larger than the charge on proteins by a factor of $\sim 2$ or more \cite{sib_rudi_ssRNA}. In physiological conditions, and when the charge on the ssRNA is about the same as the charge on the proteins (but of the opposite sign), it was found that the complexation free energy is negative so that ssRNA encapsidation takes place.

\subsubsection{Details of the energetics and conformation of ssRNA inside viral shells.~~}

A more detailed physical model of an ssRNA virus should produce the ssRNA distribution {\em as a result} of a free energy minimization. 
Different types of models of the ssRNA virus packing have been examined some of them emphasizing the 
discrete nature of the ssRNA \cite{angelescu_bruinsma}, and some, as we have seen in the previous section, representing the ssRNA 
in the continuum limit via the density field $\rho({\bf r})$ describing the spatial distribution of RNA "monomers" \cite{sib_rudi_ssRNA,schoot_bruinsma,muthukumar}. Both approaches have their advantages and limitations and they necessarily simplify the physics of ssRNA to some manageable model. Going into details of these studies is beyond the scope of this review. 

It is of interest, however, to discuss the typical spatial distributions that are obtained in these 
models. On panel b) of Fig. \ref{fig:fig_ssRNA_density} we show the ssRNA monomer density for several values of its length. The results are from Ref. \cite{sib_rudi_ssRNA}. One sees that, indeed, the ssRNA molecule occupies a shell closely separated from the capsid 
interior surface. As the ssRNA molecule becomes longer, and its charge larger from the capsid charge, the interior or the capsid 
gradually fills up. This effect is more pronounced for higher concentrations of (mono-valent) salt. The localization of ssRNA 
in a shell close to the capsid was also found in Ref. \cite{angelescu_bruinsma}, in a Brownian dynamics study of ssRNA 
(generic polymer) virus assembly where the electrostatic interactions were modeled via effective pairwise potentials \cite{Elrad_Hagan}, 
and in molecular dynamics simulation of assembled Satellite Tobacco Mosaic virus in Ref. \cite{Freddolino1}.

\begin{figure}[h]
\centering
  \includegraphics[height=11.5cm]{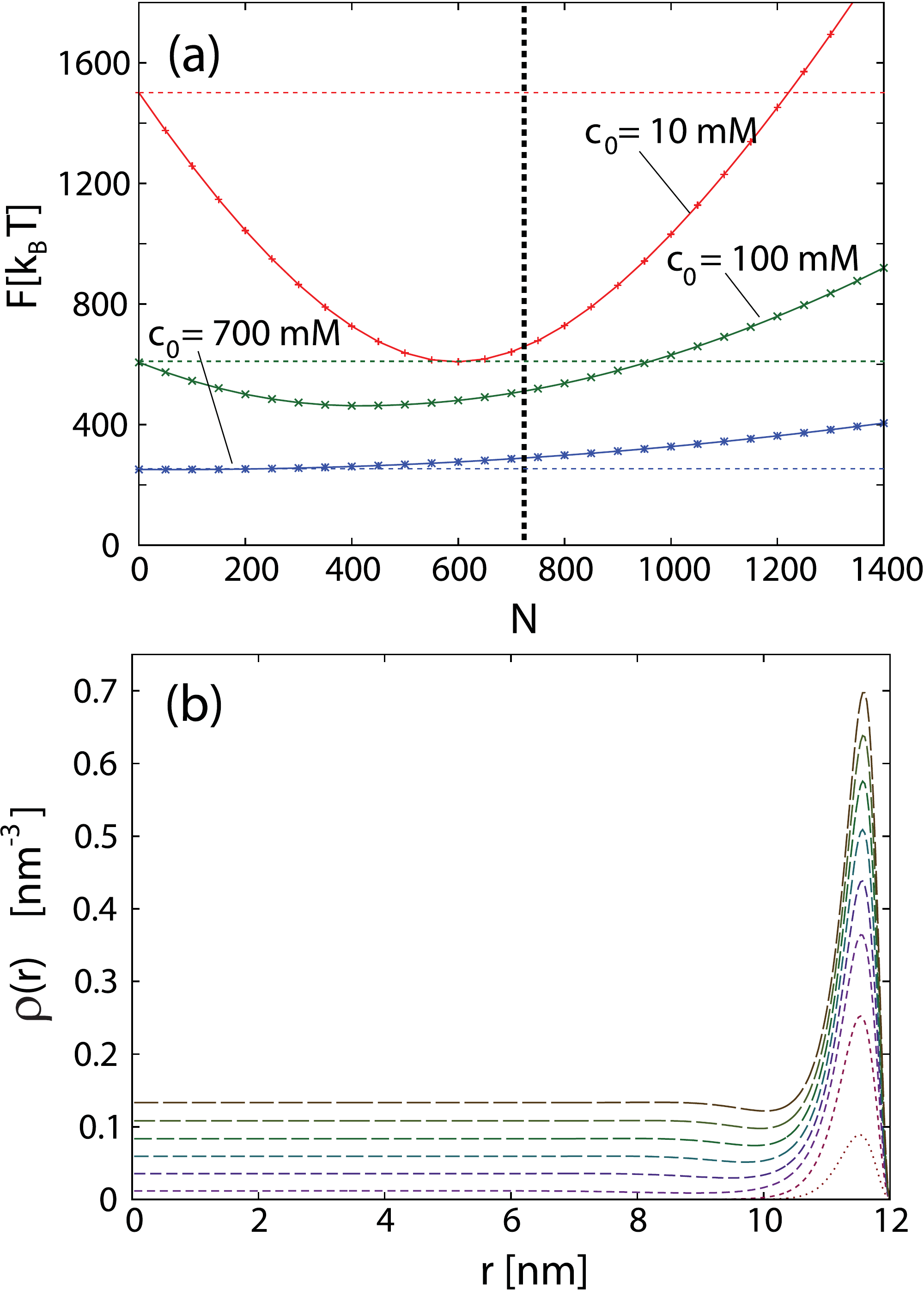}
  \caption{Panel a): The free energies of the capsid-ssRNA complexes as a function of the number of RNA bases and for three different 
monovalent salt concentration, as indicated. The free energy of the {\em empty} capsid is obtained when $N=0$. Panel b): 
The ssRNA ($a=0.5$ nm) concentration profile in a capsid of radius $R=12$ nm and surface charge density $\sigma=0.4$ $e_0$/nm$^2$ 
for $c_0$=100 mM. The curves displayed correspond to $N$ =100, 300, 500, 700, 900, 1100, 1300, and 1500 bases. The lines are 
styled so that the length of their dashes is proportional to $N$. See Ref. \cite{sib_rudi_ssRNA} for details.}
  \label{fig:fig_ssRNA_density}
\end{figure}

It was found that the energetics of ssRNA viruses also depends on the details of spatial distribution of the capsid charge, $e_0 \rho_p(r)$, in 
particular on its delocalization on the capsid protein tails that protrude into virus interior \cite{sib_rudi_ssRNA} (see 
Fig. \ref{fig:fig_virus_tails}) \footnote{Here we see the importance of the 
electrostatic interactions that is directly reflected in the shape of a virus. Furthermore, they probably underlie and influence the 
evolutionary pathway of a particular virus, whose evolutionary fitness is to a certain extent determined by the electrostatic 
interactions that govern its (dis)assembly.}.

\begin{figure}[h]
\centering
 \includegraphics[height=7.9cm]{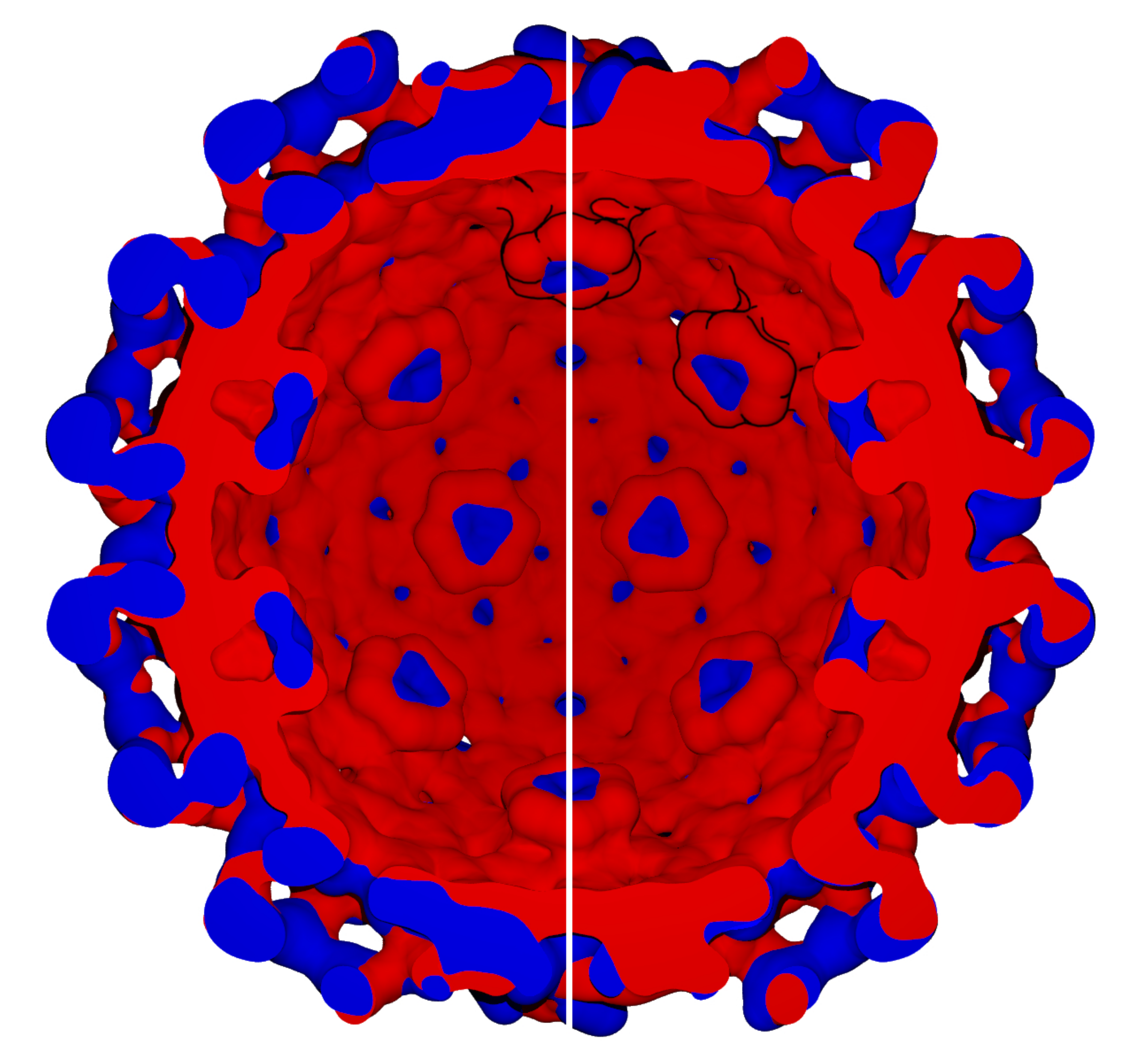}
\caption{One half of the cucumber mosaic (ssRNA) virus capsid (strain FNY). The image was constructed from RCSB Protein Databank entry 1F15 in the same manner as 
the one panel c) of Fig. \ref{fig:charges}. One can observe "buttons" of positive charge density formed by the tails of capsid proteins protruding into the interior of the virus.}
\label{fig:fig_virus_tails}
\end{figure}

The viruses that have this feature were found to be more stable with respect to changes in the ionic concentration, so that the length of the ssRNA that is encapsidated in the thermodynamically optimal conditions does not vary significantly with the salt concentration \cite{muthukumar,sib_rudi_ssRNA}. 
Figure \ref{fig:fig_tails_energies} shows the free energies of capsid-ssRNA complexes (panel a) and the profiles of the ssRNA density (panel b) obtained in the 
model that, in contrast to results shown in Fig. \ref{fig:fig_ssRNA_density}, accounts for the delocalized capsid charge, i.e. the effects of protein tails. 
The free energies do not contain the part related to attractive protein-protein interaction, so that the empty capsid electrostatic free energies (to be compared with e.g. Eq. (\ref{eq:DH_empty})) can be read out from the figures as the free energies of the complex in the limit when number of ssRNA monomers goes to zero. One sees that the optimal encapsidated ssRNA length (positions of the minima in the complex free energy curves as a function of the ssRNA length \footnote{The free energies 
at these points are estimated by Eq.(\ref{eq:DH_ss}).} in the case of infinitely thin capsid) is practically 
fixed at the position where the total ssRNA charge equals the capsid charge (vertical thick dashed line in panel a) of Fig. \ref{fig:fig_tails_energies}), almost irrespectively of the concentration of monovalent salt. One also sees that the thickness of the ssRNA shell becomes influenced by the length of the charged protein N-tails, in addition to its dependence on the screening length and monomer size (compare Figs. \ref{fig:fig_ssRNA_density} and \ref{fig:fig_tails_energies}).

The optimal encapsidated ssRNA length was found to vary much more in the case of infinitely thin capsid (panel a) of Fig. \ref{fig:fig_ssRNA_density}), so that at salinities above about 700 mM, the thermodynamically optimal ssRNA length drops down to zero and 
the model predicts formation of empty viral shells only \cite{sib_rudi_ssRNA}. In sufficiently low monovalent salt concentrations (10 mM - 100 mM), it was found 
that viruses containing the ssRNA that has about two times more charges than the capsid, are still able to self-assemble (i.e. that 
their total free energy is smaller than the free energy of assembled empty capsids and free-floating ssRNA, see Fig. \ref{fig:fig_tails_energies}). 

\begin{figure}[h]
\centering
  \includegraphics[height=11.5cm]{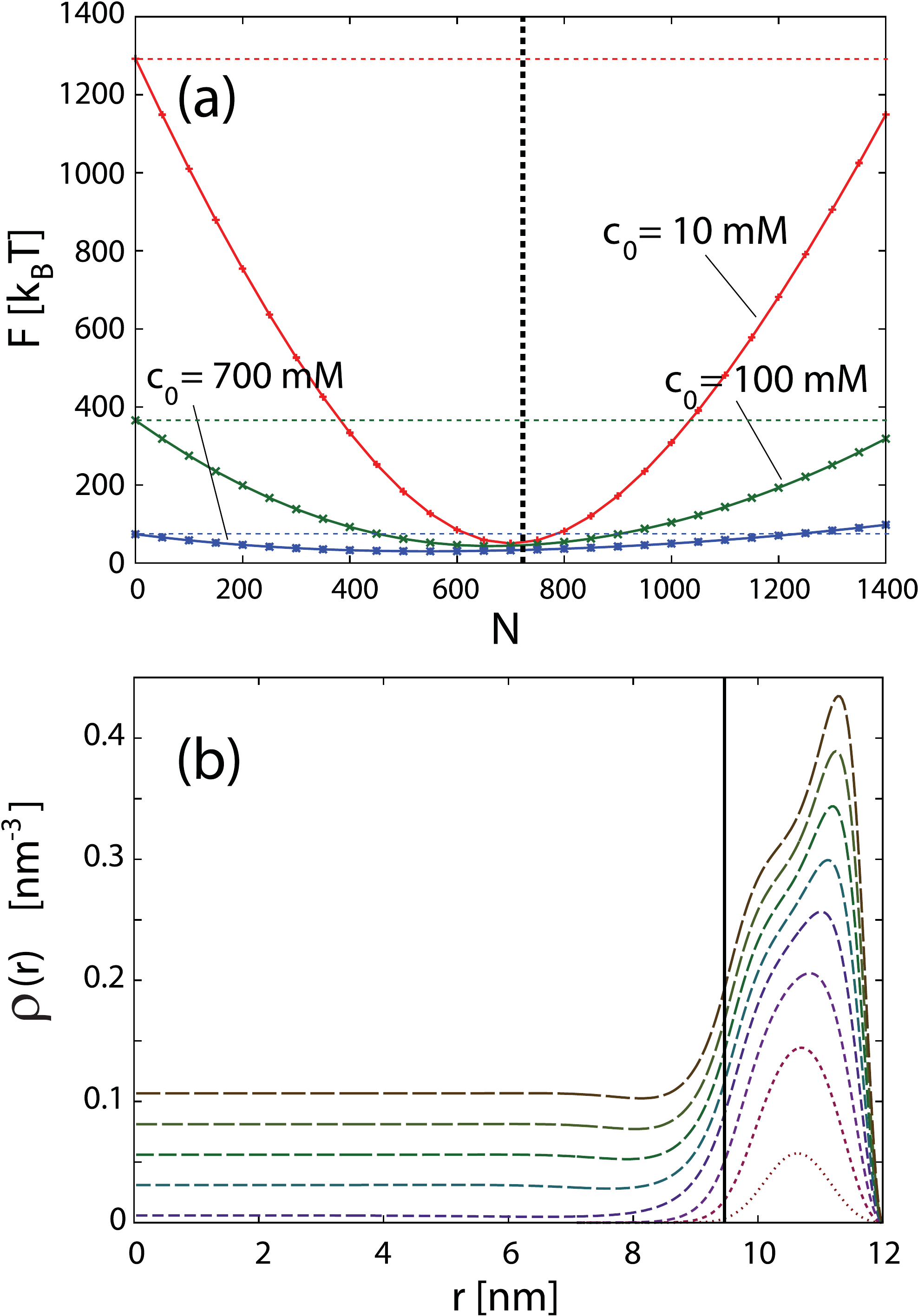}
\caption{Panel a): The free energy of the capsid-ssRNA complex as a function of the total number of ssRNA bases. The capsid radius is $R$=12 nm. The capsid 
charge is distributed in a 
spherical shell of 2.5 nm thickness. Panel b): The ssRNA density profiles for $N=$100, 300, 500, 700, 900, 1100, 1300, and 1500. 
A vertical line at $r$=9.5 nm indicates the position of the N-termini of the capsid proteins, i.e. the interior surface of the charged protein shell.}
  \label{fig:fig_tails_energies}
\end{figure}

This was confirmed in two quite different studies, one emphasizing the continuum aspects of the ssRNA \cite{sib_rudi_ssRNA} (the 
total charge on the ssRNA allowed thermodynamically and found in this study is about twice the charge on the capsid), and the other 
its discrete features \cite{angelescu_bruinsma} (there the total charge on the ssRNA allowed thermodynamically is about three times the 
charge on the capsid). This means that it is possible that the viruses carry effective charge, although the most stable state 
of an ssRNA virus is the one where the charges on the ssRNA and the capsid are about the same, i.e. when the effective charge is 
close to zero - this was found both in continuum approaches in Refs. \cite{muthukumar,sib_rudi_ssRNA} and in the approach emphasizing 
ssRNA discreteness in Ref. \cite{angelescu_bruinsma}.

All the studies discussed thus far agree that the ssRNA packages into a shell in the immediate vicinity of the inner capsid wall. There is also an interesting question of its ordering \cite{bruinsma_pressure}. Being very flexible, ssRNA is much more difficult to order even in a confined space then the much stiffer DNA \cite{Svensek,Grason}. It seems likely that the spatial inhomogeneity of the capsid charge, obeying the icosahedral symmetry of the capsid, may be responsible for ordering of the ssRNA molecule \cite{schoot_bruinsma} that has been observed in several experiments \cite{pariacoto,nodavirus}, but also in simulations in Ref. \cite{Elrad_Hagan}. It is also of interest to note that the ssRNA may locally form double stranded RNA hairpins, increasing the binding energy in this way, both via the increase of packing density and occurrence of inter-base hydrogen bonding. 

The calculations we presented do not take into account the inextensibility of ssRNA, i.e. the fact that 
the RNA monomers are linked together by chemical bonds that can not be stretched beyond their chemical size \cite{Flory}. In cases of sufficiently long and flexible ssRNA, this fact should not influence the energy significantly, however, it may be of importance for smaller ssRNA lengths. In such cases, not all monomer density distributions predicted by continuum calculations can be realized by connected monomers, and one has to explicitly account for inextensibility as was done in Ref. \cite{sib_roya_rudi}, by introducing the maximal extensibility constraint in the free energy functional.

\section{Energies of bacteriophages}
\label{sec:energies_bacteriophages}

The formation of bacteriophages in an infected bacterium proceeds by the formation of empty protein capsids first. The DNA is then inserted into a (pro)capsid \cite{principles_Cann}, an entropically and enthalpically unfavorable process \cite{michelletti3} that can be accomplished only via ATP-driven molecular motor \cite{ATPase}. 2 to 2.5 base pairs of DNA are packaged by one ATP molecule and there appear to be strong electrostatic interactions between the DNA phosphate backbone and the positively charged internal molecular motor wall \cite{molmotor}. It is not entirely clear whether there are also significant electrostatic capsid protein-DNA interaction as we have seen in ssRNA viruses \footnote{The bacteriophages do not enter the bacteria as complete particles, rather, they insert their DNA through the bacterial membrane, leaving the capsid on the outside. Strong protein-DNA interactions in bacteriophages may be disadvantageous with respect to the insertion of the DNA, i.e. they may induce ''sticking'' of the DNA to the interior of the capsid and thus prevent its entry through the bacterial membrane.} but is an experimentally observed fact that DNA molecule in the bacteriophages is nearly homogeneously distributed in their interior, so that its density is quite uniform \cite{leforestier_siber,siber_dragar}. This is an important difference in charge distribution with respect to the ssRNA viruses, and the reason for quite different energetics of the two types of viruses. The region of DNA close to the capsid shows an onion-like ordering, while the DNA in the capsid center is typically less ordered \cite{Comolli}. The outer layers most likely have a toroidal, spool-like geometry \cite{Svensek}, but there are other suggestions proposing a liquid crystal \cite{lepault}, a folded toroid \cite{Hud_Downing,Forrey}, a folded coaxial spool \cite{Serwer} and the spiral fold \cite{Black} configurations. 

The physical description of the ordered packing of the dsDNA inside the bacteriophage capsid is usually captured in one of the variants of the {\em inverse spool} model first invoked in the study of the T2 bacteriophage \cite{Kilkson,Cole,Klimenko} and formalized by Grosberg et al. \cite{Grosberg}. The inverse spool model has been subsequently refined by Odijk and Gelbart and coworkers  \cite{odijk1,odijk2,kindt,purohit1,purohit2,tzlil,klug1,klug2} and is based on the decomposition of the dsDNA total energy within the viral capsid into an interaction term and a bending term which leads to a reasonable description of the genome ejection process \cite{evil1,evil2}. Apart from the detailed simulation approaches to the DNA packing within the capsid \cite{arsuaga,spakowitz,Marenduzzo4}, all theoretical work is based on assumptions regarding the form of the curvature energy of the DNA forced to reside within the confines of the capsid, as well as the interactions among the highly charged and hydrated DNA segments  packed at high densities within the capsid. 

Elastic curvature energy appears to be the lesser of the two unknowns, though some very recent work might point to the contrary \cite{nelson}. It is proportional to the square of local DNA curvature and in fact follows from the Euler-Kirchhoff model of an elastic filament. Though this model contains some subtle features due to the strong interhelical forces between the segments of the molecule \cite{Rudi_elastic}, it nevertheless appears to be a consistent description of DNA \cite{rouzina} on mesoscopic scales \cite{nelson}. The parameters of the Euler-Kirchhoffian model of DNA, such as its persistence length, are well established and have been measured by a variety of methods with satisfactory consensus among the results \cite{hagerman_flexibility}.

\subsection{Electrostatic interactions and energies of bacteriophages}

As the DNA molecule is highly charged (two elementary charges per base pair), its insertion in the bacteriophage interior results in a buildup of repulsive electrostatic force and energy \cite{Bustamante}. The electrostatic energy of a rough model where the DNA molecule is homogeneously distributed in the capsid interior can be estimated in the DH framework using a technique similar to that presented for derivation of Eq. (\ref{eq:DH_empty}). In the limit $\kappa R \gg 1$, one 
obtains 
\begin{equation}
\lim_{\kappa R \gg 1} F_{DH} = \frac{2 \rho^2 R^3 \pi}{3 \kappa ^2 \varepsilon_0 \varepsilon},
\label{eq:bacteriophage_energy}
\end{equation}
where $\rho$ is the volume density of the DNA charge (2 elementary negative charges per base pair). The electrostatic energy can now 
be estimated in the case of $\lambda$ bacteriophage, whose radius is $R \approx 30$ nm, and whose DNA contains $N_{BP}$=41500 base pairs. 
The density is $\rho \sim 0.7$ $e_0$/nm$^3$, and the electrostatic energy in physiological conditions $F_{DH} \sim 3$ 10$^5$ 
$k_B T$. This is at least an order of magnitude larger from the repulsive electrostatic energy of ssRNA viruses, although the 
protein-DNA interaction in bacteriophages may reduce this energy somewhat (the attractive protein-DNA contribution scales 
with $R^2$, however, so it becomes less relevant for larger bacteriophages). The total free energy balance upon assembly can 
hardly be made negative by the attractive protein-protein interactions. It then follows that the bacteriophages can not self assemble 
\footnote{The attractive energy of protein-protein interactions scales with the total length of 
protein contacts (buried surface area), i.e. with $R^2$. In case of ssRNA viruses, the repulsive electrostatic 
energy, for given surface charge density scales also with $R^2$, which suggests that ssRNA viruses of similar surface charge 
densities but different radii (i.e. made of similar proteins, but of different $T$ numbers) can all rely on attractive 
protein-protein interactions in order to self assemble. This is very unlike the case of bacteriophages where the electrostatic 
energy of DNA, for given packing density scales with $R^3$.} and they must therefore use other means of assembly. 

Electrostatics accounts for only a part of the interaction energy between DNA segments. In fact these interactions can be measured directly in osmotic stress experiments \cite{methods} and can be deconvoluted into a longer ranged electrostatic contribution \cite{rudi-rev-int} and a shorter ranged hydration component \cite{leikin}. Both of them have been quantified in terms of magnitudes and decay lengths \cite{Podgornik3}. The various formulations of the inverse spool model mostly differ in terms of the exact form of the interaction potential. While some are based entirely on theoretical polyelectrolyte models \cite{slok,odijk1,odijk2}, others are based on semi-empirical chemical potential expressions \cite{purohit1,purohit2,klug1,klug2}. The best strategy would be of course to use directly the measured osmotic pressure from the osmotic stress experiments as an input for the formulation of the theory \cite{siber_dragar}.

\subsection{Elastic energy of DNA packing}

A simple calculation we used to estimate the electrostatic energy of the bacteriophage DNA, Eq. (\ref{eq:bacteriophage_energy}), does not 
take into account the intrinsic stiffness of the dsDNA and the estimate obtained would be the same were the DNA base pairs completely disconnected from one another, i.e. packed homogeneously within the capsid as a cloud of charged monomers. The fact that DNA is confined and thus substantially coiled brings its bending energy clearly into focus. The bending elasticity of DNA can be accounted for within the Euler-Kirchhoff elastic energy \cite{Marko_Siggia,neukirch} and introduces an additional length scale, ${\cal L}_P$, to the problem, referred to as the persistence length \cite{hagerman_flexibility}. The persistence length represents the length scale over which the direction of the DNA  is correlated and was measured to be about 50 nm in physiological conditions \cite{Bustamante_persistence,Peterlin}. One should note here that this number includes also an electrostatic contribution to the DNA elasticity, i.e. the fact that charges on an elastic string also contribute 
to its elasticity \cite{Bustamante_persistence,roya_persistence}, as they want to remain as distant as possible, favoring thus flat DNA strand conformations. One can describe this as an electrostatic renormalization of the ''bare'' (chemical) DNA elasticity, i.e. the elasticity that the DNA strand would have were its bases uncharged \cite{odijk,skolnick,Rudi_elastic}. The bare persistence length, ${\cal L}_P$ is thus smaller than 50 nm. 

To estimate the elastic contribution to the DNA packing, one can use techniques similar to those presented in Refs. \cite{UO,Purohit}. The 
elastic energy of packing can be expressed in different equivalent continuum forms \cite{siber_dragar}. In what follows we use the form 
\begin{equation}
U_{elastic} = \frac{{\cal L}_P L_{BP}}{2} k_B T \int d^3 r \frac{\rho_{BP}(r)}{R^2({r})},
\label{eq:elastic_UO}
\end{equation}
where ${\cal L}_P$ is the persistence length of DNA, $L_{BP}=0.34$ nm is the spacing between the DNA base pairs, $\rho_{BP}(r)$ is the DNA base-pair density (the charge density is thus $\rho(r) = 2e_0 \rho_{BP}(r)$) at a radial distance ${r}$ from the axis of DNA packing,  and $R^2({r})$ is the squared radius of curvature of DNA, the assumption being that the configuration of DNA orientational order has cylindrical symmetry. 

The elasticity of DNA should induce an inhomogeneity in the DNA packing density, $\rho_{BP}(r)$, which would be constant were the electrostatic energy the only part of the 
total free energy. Looking at Eq. \ref{eq:elastic_UO}, one sees that the elastic contribution diverges as the radii of curvature 
approach to zero. Assuming that the DNA is packed in a bacteriophage capsids in a spool-like manner, the regions 
which are most disfavored with respect to elastic energy are those close to the spooling axis, where the radius of curvature, 
$R$, is small, decreasing to zero on the axis. A reasonable guess of $\rho(r)$ (or variational ansatz) that includes the 
effects of elasticity would be a homogeneous density of DNA outside a cylinder of radius $R_0$, that drops to zero inside 
the cylindrical void, i.e. the zone of DNA exclusion due to elastic effects \cite{Purohit,Odijk_bacteriophage}. This approach is amenable to analysis, as it 
allows one to combine the elastic and electrostatic energies of a DNA distribution with exclusion radius $R_0$, and to 
find the radius, requiring that the total energy be minimal. The result in the case when $\kappa R \gg 1$, $R_0 \ll R$, 
$N_{BP} \gg 1$, $R = 30$ nm (bacteriophage $\lambda$), and ${\cal L}_P = 50$ nm (maximal estimate for bare DNA persistence length) 
is \cite{sib_anze_rudi_unpub1}
\begin{equation}
R_0 \sim \frac{10 R}{\sqrt{N_{BP}}}.
\end{equation}
This equation indicates small effects of elasticity on the distribution of packed DNA. For fully packed bacteriophage $\lambda$, 
$N_{BP}=41500$, and the exclusion radius amounts to $R_0 \approx 0.05 R$. This is smaller (1.5 nm) than the diameter of the 
DNA strand ($\sim$ 2.5 nm) which means that the exclusion void, if it exist, is determined 
by the discrete nature of the DNA molecule, and the minimum possible length of a DNA kink. Note also that the value of the persistence 
length that we used in the calculations above (50 nm) includes some of the effects of electrostatic interactions that have already 
been included in Eq. (\ref{eq:bacteriophage_energy}), so that the void radius is expected to be even smaller.

It is also possible that nonlinear elastic effects and knotting \cite{harvey-virus,Marenduzzo3} may be of importance for such a highly bent DNA configuration \cite{Wiggins}. The relative contribution of elastic energy to the total energy of packing in this model scales with $N_{BP}$ as $\ln N_{BP}/N_{BP}$. In the case of fully packed bacteriophage, it amounts to less than 5 \% of total energy \cite{sib_anze_rudi_unpub1}. A similar estimate has also been obtained on the basis of experimental data in 
Ref. \cite{parsegian_bacteriophage}. Smallness of elastic contribution to the total energy has been discussed in Ref. \cite{siber_dragar}, using different techniques that do not attempt to estimate the electrostatic contribution to the DNA packing but rather utilize the data from bulk DNA osmotic pressure experiments \cite{rudi-rev-int} in order to obtain the required information. The region of depleted DNA density is not seen in experiments \cite{leforestier_siber,bustamante_bacteriophage}. Rather, a more disordered phase of DNA is seen around the center of the capsid, which does not have an ''onion-like'' type ordering characteristic for DNA packed close to the capsid interior.

The contribution of DNA elasticity to bacteriophage packing energy and the inhomogeneities in the density of the encapsidated DNA are both quite small when the DNA is in a non-condensed state \cite{siber_dragar,petrov}, i.e. when the interactions between the DNA segments are repulsive. In fact this is always the case either when the counterions in the bathing solution are monovalent and/or there are no crowding agents - such as poly-ethylene-glycol - present in the bathing solution. Monovalent counterions can change the strength of intersegment DNA interactions but they can not change their character, i.e. they can not turn repulsive interactions into attractive interactions \cite{rudi-rev-int}. 

Monovalent salt ions in general renormalize the bare phosphate backbone charge along DNA as well as determine the range of electrostatic interaction via their screening properties. The electrostatics of the DNA \cite{Cherstvy,Kornyshev-rev} becomes complicated in the presence of strongly charged counterions, basic proteins, crowding agents, the DNA confinement, etc. \cite{michelletti4} . One of the salient features of these complicated interactions is that adding even small amounts of polyvalent counterions (spermine, spermidine, CoHex, ...) changes the nature of the electrostatic interactions so that they can turn attractive even between nominally equally charged DNA segments where one would expect repulsions \cite{parsegian_bacteriophage,rau_parsegian,rau_parsegian2}. The same effect can be observed also in
the bacteriophage with a partially ejected genome \cite{leforestier_siber} where addition of spermine {\em condenses} the remaining DNA into a toroidal spool completely contained within the bacteriophage capsid. In this case the contribution of the elastic energy to the equilibrium free energy is non-negligible being in fact essential in order to counteract the attractive interactions that would tend to compact DNA into a disordered globule.

\section{Osmotic pressure in viruses}
\label{sec:pressures}

Since solvent can equilibrate across the viral capsid the mechanical pressure acting on the capsid wall equals the osmotic pressure of the solution enclosed within the capsid \cite{Panja,Molineux}. This osmotic pressure has various contributions that differ in the case of flexible ssRNA viruses as opposed to the stiff dsDNA bacteriophages due to the nature of the polymer elasticity. In the former case we have purely entropic elasticity stemming simply from the connectivity of a flexible chain, while in the latter case we have enthalpic elasticity of a semiflexible Eulerian filament very different in form and magnitude \cite{Grosberg}.

\begin{figure*}
\centering
  \includegraphics[width=14cm]{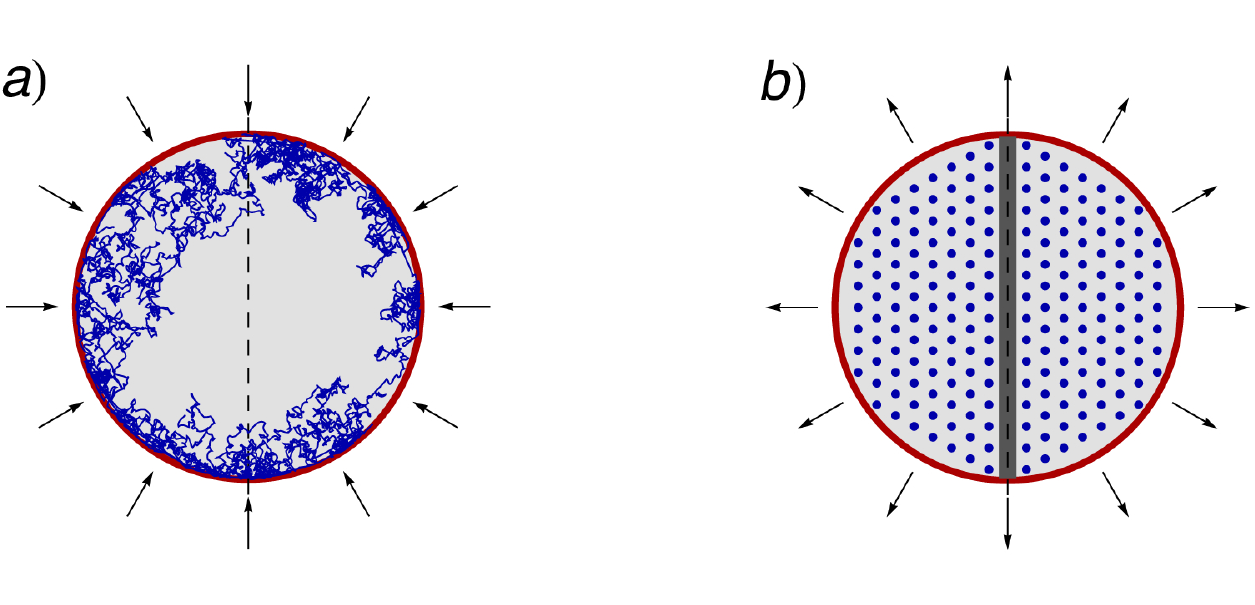}
  \caption{Pressures sketch for an ssRNA [Panel a)] vs dsDNA virus [Panel b)]. Genome is shown in blue, capsid in red, arrows indicate the direction of pressure. Cross section through the capsid. In the ssRNA case the genome is disordered but shows a distinct boundary layer close to the hypotope of the virus with a depleted region in the middle. In the dsDNA case the genome density within the capsid is (almost) uniform, showing pronounced orientational and positional ordering.}
  \label{fig:pressv1}
\end{figure*}

The differences in molecular flexibility of the encapsidated nucleic acid component furthermore engender also qualitative differences in the ordering of the genome. In the case of ssRNA its flexibility implies, in the simplest case, a completely disordered polymer solution of homogeneous density as can be observed experimentally with e.g. hyaluronic acid \cite{Reed}, while in the case of dsDNA at high concentrations its stiffness leads to an orientationally and positionally ordered mesophase \cite{mesophases,Coll_DNA}. This idealized dichotomy ceases to be valid in the context of viruses at special conditions when ssRNA can assume a more ordered hypotope-bound configuration reflecting the icosahedral symmetry of the inner capsid wall \cite{bruinsma_pressure}. The packing of dsDNA within bacteriophages seems to be governed by the stiffness of DNA that leads to the formation of local nematic alignments \cite{lepault} whose orientational order is constrained by the bacteriophage capsid. The orientational ordering of dsDNA within the bacteriophage can thus be seen as a type of constrained liquid crystalline ordering that can be analyzed by a local thermodynamic mesoscopic theory \cite{kulic,Svensek,Grason}. 

In a thermodynamic equilibrium osmotic pressure at every point within the capsid has to be the same and can be set by external osmoticants such as PEG \cite{evil3,cohen}. For various models of the viral core comprised of the nucleic acid component and the bathing ionic solution, the total osmotic pressure can be decomposed into separate terms stemming from the different components of the bathing solution. Within the mean-field approximation of the ionic component of the bathing solution it follows that its osmotic pressure is given by the van't Hoff ideal osmotic pressure of all the ionic components \cite{Andelman_pap} at the hypotopic wall of the capsid. As for the nucleic acid component it is given by an additive term due to either the entropic elasticity of a flexible chain  \cite{rudi_polyel} or enthalpic elasticity of a semiflexible chain \cite{siber_dragar}.

Specifically, the case of  flexible polyelectrolyte chain(s) such as ssRNA in an ionic solution the total osmotic pressure decouples into two contributions \cite{rudi_polyel}:
the first one is the osmotic pressure of ions inside the capsid due to electrostatic interactions between the ions themselves, as well as between the ions and the polyelectrolyte and the capsid charges; within the PB frame it equals the ideal van't Hoff expression for all the components of the solution, except the polyelectrolyte chain, evaluated at the inner surface of the capsid wall; the second one takes into account the connectedness of the polyelectrolyte chain and its interaction with the mean electrostatic field. While the first term is positive, the second one can be of either sign. When it becomes negative we refer to it as the {\em polyelectrolyte bridging} contribution \cite{rudi-PE}. Neither of the two terms is in general dominant and the overall sign is due to a subtle interplay of electrolyte and polyelectrolyte properties of multicomponent solution within the virus capsid \cite{rudi_polyel,rudi_poly_2,rudi-PE}.

For the stiff dsDNA polyelectrolyte a different decomposition can be derived \cite{siber_dragar}. In that case the osmotic pressure again decouples into two terms:
the first term now corresponds to all the self-interactions between the segments of the encapsidated DNA molecule \cite{rudi-rev-int}, whereas the second one is the contribution of the elastic bending deformation of the ordered DNA mesophase imposed by the capsid confinement \cite{Svensek,Grayson}. While the second term is strictly positive, since any deformation of the orientation of the DNA molecule increases its free energy, the first one can be of either sign depending on the nature of interactions between DNA molecules \cite{rudi-rev-int}.  In principle it contains {\em all} interactions between DNA molecules that could be either electrostatic or non-electrostatic in nature. Comparing the measured interaction term with the curvature term one can make a general conclusion that the latter is important only in the immediate vicinity of the central axis of DNA packing symmetry, consistent with the model calculations of bacteriophage energy presented in the previous section.

We now estimate the magnitude of the osmotic pressure in both types of viruses by making a very restrictive assumption that the interactions are strictly electrostatic in origin. This allows us to use simple scaling arguments in the derivation.

\subsection{Osmotic pressure in self-assembled ssRNA viruses}

The scaling form of the polyelectrolyte osmotic pressure, $p_{PE} $, acting in ssRNA viruses can be obtained from equation Eq. (\ref{eq:DH_ss}) if one assumes that electrostatic interactions can be described on the DH level and that confined ssRNA distribution can be described as a dense hypotopal layer of thickness $a$ and a depleted core. One first needs to express the equation for the virus free energy Eq. \ref{eq:DH_ss} within the DH approximation in terms of the total charge on the capsid ($Q$) and on the ssRNA ($\sim -Q$), 
\begin{equation}
F_{DH} \approx F_{DH}(-\sigma, \sigma, a) = \frac{f Q^2 a}{16 R^2 \pi \varepsilon_0 \varepsilon}.
\end{equation}
The osmotic pressure of the polyelectrolyte (PE) chain can then be calculated from the appropriate derivative of the free energy at constant number of surface charges, 
\begin{equation}
p_{PE} = - \frac{1}{4 R^2 \pi} \left. \frac{\partial F_{DH}}{\partial R} \right |_{Q}.
\end{equation}
The sign here corresponds to the standard definition with pressure being positive, i.e. repulsive, for free energy that decays with radius. This gives
\begin{equation}
p_{PE} \approx \frac{f Q^2 a}{32 R^5 \pi^2 \varepsilon_0 \varepsilon}.
\end{equation}
For a "typical" ssRNA virus, this gives $p_{PE} \sim 10$ atm, acting to increase the capsid radius (outward). There is, however a component of pressure missing in the above evaluation. This is the component related to attractive protein-protein interactions  in the capsid. 

Namely, there is an outward pressure acting in the {\em empty capsid} also, since in this case the charges on the capsid decrease their energy by separating. On the DH level this purely electrostatic (ES) component of osmotic pressure is then given by
\begin{eqnarray}
p_{ES}^{empty} &=& -\frac{1}{4 R^2 \pi} \left. \frac{\partial F_{DH}^{empty}}{\partial R} \right |_{Q} = - \frac{Q^2}{32 R^5 \pi^2 \varepsilon_0 \varepsilon \kappa},
\end{eqnarray}
where $F_{DH}^{empty}$ is the free energy of an empty capsid given in Eq. (\ref{eq:DH_empty}). If we assume that the empty capsid of an ssRNA virus in question can self-assemble, we 
are led to propose that the repulsive electrostatic pressure in {\em empty capsids} is exactly, or to a good proportion canceled by 
pressure arising from the attractive interactions. %
Were the proteins infinitesimally small particles, the two pressures should exactly cancel in the assembled empty capsid. 

If we include the attractive protein-protein interactions in the calculation of total pressure, it is then given by
\begin{equation}
p = p_{PE} + p_{ES}  \approx \frac{Q^2}{32 R^5 \pi^2 \varepsilon_0 \varepsilon} \left( fa - \kappa^{-1} \right).
\label{eq:press_ssRNA}
\end{equation}
Since $f a \sim 1$ nm, and $\kappa^{-1} \sim 1$ nm, we see that the total pressure in ssRNA virus {\em almost vanishes}, since the 
different contributions nearly perfectly cancel.

\begin{figure}[h]
\centering
  \includegraphics[width=9.5cm]{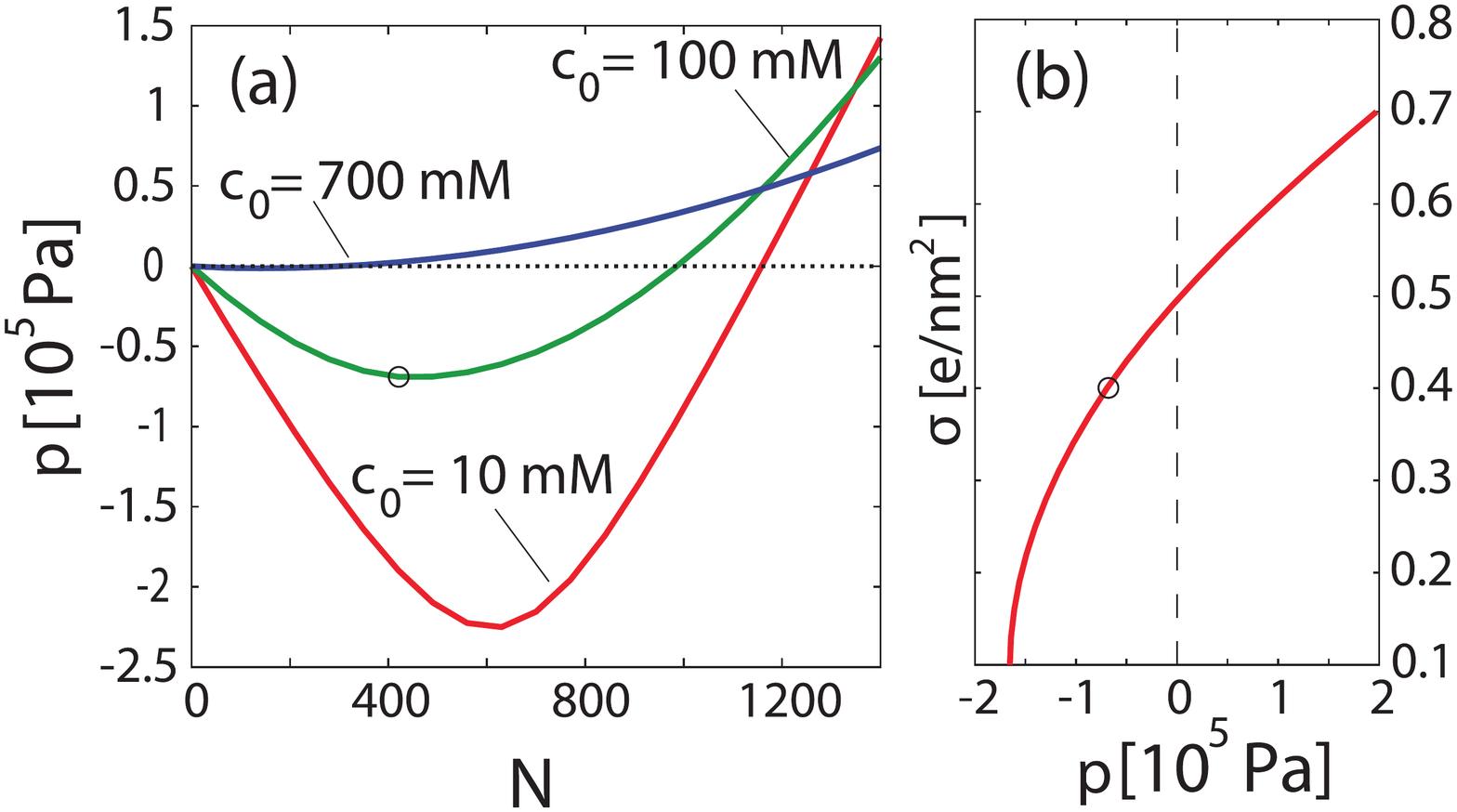}
\caption{Panel a): Osmotic pressure acting on the viral capsid as a function of the number of monomers. Panel b): Osmotic pressure for fixed number of monomers $N=420$, $c_0$= 100 mM
[the point denoted by a circle in panel (a)] as a function of the capsid charge density, $\sigma$. The capsid is assumed to be infinitely thin, with radius of $R$=12 nm. Other parameters 
of the calculation are the same as in Fig. \ref{fig:fig_ssRNA_density} (adapted from Ref. \cite{sib_rudi_ssRNA}).}
  \label{osm_pres}
\end{figure}

In more detailed calculations of osmotic pressure \cite{sib_rudi_ssRNA}, it was found that the pressure in ssRNA virus can be either positive or negative, depending on the total charge on the virus and its relation to other parameters in the problem, see Fig. \ref{osm_pres}, but it was found to be slightly negative (half an atmosphere) at the point of optimal assembly, i.e. at the point where the length of the ssRNA is such to minimize the complexation free energy functional. 

The negative values of osmotic pressure can be understood by analogy with charged planar surfaces with an oppositely charged polyelectrolyte chain in between and are due to bridging configurations of the chain \cite{rudi-PE}. One could then connect the negative values of the osmotic pressure in the case discussed here, with a similar mechanism. The ssRNA enclosed within the capsid would thus bridge the space between neighboring sections along the curved capsid and induce attractive interactions between them. Summing these attractions along the total surface of the capsid would give rise to an overall negative osmotic pressure of the polyelectrolyte. This interpretation is certainly corroborated by the characteristic density profile of the polyelectrolyte concentration in the vicinity of the capsid wall, which is in fact very similar to the bimodal polyelectrolyte concentration profile observed in the case of planar polyelectrolyte confinement \cite{rudi_polyel,rudi_poly_2}.

\subsection{Osmotic pressure in bacteriophages}

In the case of the dsDNA bacteriophage the osmotic pressure due to DNA self-interaction is in general much larger than the one due to the DNA elasticity, which we can thus safely ignore. Assuming furthermore, that the only interaction between DNA molecules is electrostatic we can estimate its magnitude with scaling arguments. There is nevertheless large variation in the magnitude of the osmotic pressure for various virus types even if it is mostly of electrostatic nature \cite{Fuller}.

Osmotic pressure arising from the dense packing of DNA in bacteriophages can be obtained by examining how the electrostatic energy of encapsidated  DNA changes upon the increase of the capsid radius. The energy should be expressed in terms of the total number of DNA base pairs which is conserved in the process of infinitesimal increase of the radius, i.e.
\begin{equation}
p \approx - \frac{1}{4 R^2 \pi} \left. \frac{\partial F_{DH}}{\partial R} \right | _{N_{BP}}.
\end{equation}
Using Eq. (\ref{eq:bacteriophage_energy}) in the limit of $\kappa R \gg 1$ this yields
\begin{equation}
p = \frac{9 N_{BP}^2 e_0^2}{8 \pi^2 R^6 \kappa^2 \varepsilon_0 \varepsilon}, 
\label{eq:press_DNA}
\end{equation}
which to the lowest order coincides with the estimates based on the Donnan potential \cite{slok,odijk2,deserno}. Osmotic pressure is thus positive corresponding to a repulsive force acting on the capsid wall trying to increase its radius. The above equation can be evaluated for bacteriophage $\lambda$ and it gives $p \sim$ 100 atm, which is somewhat higher than measured in experiments \cite{Bustamante}. On one hand, this is the pressure that the bacteriophage capsid needs to withstand, and on the other this is the pressure of the DNA coiled osmotic spring piled up against the inner surface of the capsid ready to release its chemical and mechanical energy through the portal complex on docking onto a bacterial wall \cite{phages}.

For real bacteriophages at a dsDNA concentration corresponding to $\sim 2.7$ nm inter DNA spacing, which amounts to $\sim \rm ~500 mg/ml$, the osmotic pressure is given not only by the electrostatic interaction between DNA molecules \cite{inspired} but also by the hydration interactions stemming from the ordered vicinal layers of water close to the DNA surface \cite{rudi-rev-int}.

\subsection{Osmotic pressure: ssRNA viruses vs. dsDNA bacteriophages}

Osmotic pressure of the two types of viruses  \cite{evil3} differs crucially in its magnitude as well as sign. While for ssRNA it is usually small and could be negative, it is strictly positive and large for bacteriophages. Negative osmotic pressures are ubiquitous in polymer shells that are impenetrable to osmoticants \cite{Brumen,Brumen2,Vinogradova}. The difference of osmotic pressures also mirrors the very different assembly paths of the two types of viruses: spontaneous self-assembly  at almost vanishing osmotic pressure difference vs. non-equilibrium (active) packing mechanisms where DNA encapsidation has to fight an enormous positive osmotic pressure inside the capsid \cite{evil4}. These salient features of osmotic pressure carry on even into more realistic models of interactions between the nucleic acid component of the viral core \cite{rudi-rev-int}. 

The scaling forms of electrostatic interaction contribution to osmotic pressure of ssRNA and dsDNA viruses also convey another interesting contrast between the two. While in the case of the dsDNA bacteriophages the electrostatic interaction energy stems largely from the interactions acting between the segments of nucleic acid component within the volume of the virus \cite{podgornikreview}, in the case of the ssRNA viruses the most important contribution comes from the interaction between the ssRNA genome and the hypotopic surface of the capsid. This contrast in the interaction mechanisms shows up eventually as the $R^{-6}$ scaling for DNA viruses vs. the $R^{-5}$ scaling for the ssRNA viruses in the simple picture introduced above. 

\section{Electrostatic effects in viruses for highly charged counterions}
\label{sec:strong}

Many bacteriophages require Mg$^{2+}$, Zn$^{2+}$ and Ca$^{2+}$ ions in order to attach to bacterial membrane \cite{principles_Cann}. 
Spermine ($\rm Spm^{4+}$) and other multiply charged polyamines have been reported in a variety of viruses \cite{Carter}. Arginine rich protamines are essential for sperm head condensation and DNA stabilization in spermatogenesis \cite{protamins,protaminsjason} and the effect of long polycations such as poly-L-lysine, poly-L-arginine and linear and branched polyethylene imine on interactions between DNA molecules have been studied in detail \cite{jason-netz,jason-adrian}. 

The electrostatic interactions mediated by the polyvalent counterions can not in general be discussed within the mean-field {\em ansatz} and their presence has consequences that do not have their counterpart in the monovalent case \cite{inspired}. We now understand that polyvalent counterions even in mM concentrations mediate a different type of electrostatic interactions than the monovalent counterions \cite{naji-rev,hoda-rev}. They can drastically modify either interactions between the nucleic acid component of the virion, or the interaction between the capsomere subunits of the capsid. In both cases their effect could most probably be ascribed to a diminished repulsion or indeed net attraction between the similarly charged macromolecular subunits\cite{Qiu}. In the context of virus electrostatics these effects have not been analyzed yet and can be of several types that we address separately.

\subsection{Specific binding of polyvalent counterions}

Ions even of the same net charge have many properties which are specific for each ion type \cite{Ben-Yaakov2}. It is well known that their interaction with the DNA surface is not governed by the charge only but is a complicated combination of their size, hydration and polarization properties \cite{plavec}. Ion specific effects are quite important \cite{rau_parsegian,rau_parsegian2} but are notoriously difficult to understand \cite{Kunz,Kunze}. Specifically in the context of DNA interactions the nature of the ions sets the magnitude of the repulsive electrostatic interaction in the case of monovalent counterions \cite{Parametrization}, as well as the emergence of counterion mediated attractions leading to the onset of DNA condensation \cite{Rau_rev}. At present the best one can do is to include the ion specific effects on some kind of a phenomenological level as is the case when dealing with counterion adsorption along DNA \cite{Kornyshev-rev}. 

Ionic specificity becomes particularly important when analyzing the details of the DNA ejection mechanisms \cite{evil5} or capsid protein assembly and stability \cite{evil6}.  Capsid proteins often contain Angstrom-sized ''voids'', holes, i.e. three-dimensional empty spaces which act as geometrical sieves, enabling entry of 
ions of small enough size \cite{Ptitsyn}. When the void is surrounded with a particular distribution of protein charges, the void may act also as 
an electrostatic sieve, permitting entry and localization of only sufficiently charged ions of proper effective size. In such a 
localized configuration of counterions and the protein, even quantum effects may be important, which may further differentiate between
specific electronic structure of different ions. All this may result in a specificity of counterion binding. Note that the mechanism discussed 
gives rise to a {\em localization} of the counterions, an effect that is not included in the PB and DH approaches. This results 
in effective {\em binding} of the counterion in the holes formed by charged protein groups \cite{Oosawa_book}, an effect that  in 
the context of protein interactions is usually known as {\em coordinate bonding} and involves di- and trivalent ions of Fe, Zn, Co, 
Ca, Mg, as well as other metallic ions \cite{Ptitsyn}. 

The role of such ions may be important both for establishment of the capsid protein fold  (i.e. the functional configuration of the 
capsid protein or proteins in general \cite{garciamoreno}), and also for the binding of proteins to form a capsid. In this respect, the presence of these ions may act as a switch, yielding an effective attraction between the proteins when the (multivalent) specific counterions are added in required (small) concentrations. Attractions of this type can be invoked also in other contexts of the nano-scale interactions \cite{French_RMP}. Thus, these effects that can be quite difficult to quantify can be considered as contributing to the attractive component of the protein-protein interactions, and the approach presented still serves to establish the importance of monovalent ions in the repulsive electrostatic interactions between the proteins.

\subsection{``Dressed counterion'' approximation: nonspecific effects of polyvalent counterions}

As we already stated electrostatic interactions in the presence of polyvalent counterions or highly charged macroions can not be properly understood on the mean-field level. Sometimes even the sign of the interaction predicted by the mean-field theory is wrong. While this shortcoming of the PB theory has been known in the general colloidal context since the mid-eighties \cite{Gulbrand,LeBret}, it made its debut in the study of DNA interaction only about 10 years later \cite{Niels,Lars}. The salient features, not the ion specific effects, of the polyvalent counterion mediated interactions can be formulated analytically based on the {\em strong coupling} electrostatics that has been pioneered by Rouzina and Bloomfield \cite{Rouzina-Blo}, elaborated later by Shklovskii et al. \cite{Shklovskii_colloq}, Levin et al. \cite{elablevin}, and brought into final form by Netz et al. \cite{Jungblut,hoda-rev}. An important feature of the strong coupling regime is that the counterion mediated interactions between nominally equally charged macroions can become attractive and can indeed cause condensation of DNA \cite{Khan,Svensek2}. 

As the bathing solution of viruses often requires the presence of polyvalent counterions, the question remains of the nature of the polyvalent counterion effect in the electrostatics of viruses, interactions between DNAs notwithstanding. We are {\em not} addressing the specific binding effects \cite{Kornyshev-rev,LeeJCP} but the {\em non-specific} universal features of the polyvalent counterions. These effects have not been analyzed in detail yet, but there are some features that can be addressed without delving too deeply into the theory of strongly charged Coulomb systems. Usually, the aqueous solution of viruses contains a mixture of monovalent as well as polyvalent salts at relatively high and low concentrations, respectively. An intermediate approximation termed the {\em dressed counterion} (DC) approximation allows for an approximate treatment of all the components in this highly asymmetric multicomponent electrolyte system \cite{dressed1,dressed2}. It starts with the observation that the concentration of polyvalent counterions is usually small and thus can be dealt with on the lowest order virial expansion level, whereas the monovalent salt ions can still be treated on the mean-field DH level.

Within the DC approximation the electrostatic free energy of a charged shell is composed of two parts: the first one is the DH free energy that we already evaluated before, Eq. (\ref{eq:DH_empty}), and is due to the electrostatic interaction between the charges on the capsid and the monovalent salt ions of the bathing ionic solution. The second part is due to the presence of a small concentration of polyvalent counterions interacting via the DH screened potential. In the limit of small concentrations and highly charged counterions the contribution of the polyvalent counterions can be well approximated by the first order virial expansion corresponding to a single counterion interacting with all the charges in the system, i.e. the monovalent salt ion charges as well as the fixed charges on the capsid \cite{dressed1,dressed2}. The grand canonical potential can thus be written
\begin{eqnarray}
\beta\Phi&=&\beta F_{DH}-\beta F_{DC}^{(1)}=\nonumber\\
&=&\beta F_{DH}-n_{DC}\int_V\exp\left(-\beta e_0 v\phi_{DH}(\mathbf{r})\right)\,\mathrm{d}^3\mathbf{r},
\label{eq:free_DC}
\end{eqnarray}
where $n_{DC}$ is the bulk concentration of the added $v$-valent conterions. In a typical case of virus assembly, the solution contains relatively large concentration of monovalent salt ($\sim$ 100 mM) and by comparison a small concentration of $v$-valent ions ($\sim$ 1 mM). The electrostatic potential $\phi_{DH}$ is again the solution of the Debye-H\"uckel equation for the given system.

The DC part of the total electrostatic free energy contribution $F_{DC}^{(1)}$ scales quite differently with the capsid surface charge density, $\sigma$, and the DH screening length, $\kappa$, than the mean-field contribution $F_{DH}$ [Eq.~(\ref{eq:DH_empty})]. This is also the case when the DC calculation is performed for a sphere homogeneously filled with charge, which is a simple model of a bacteriophage [with the free energy in the DH limit given by Eq.~(\ref{eq:bacteriophage_energy})]. Since the $v$-valent ions and the bulk/surface charges are usually of opposite signs, the DC part of the total free energy acts to lower the DH contribution, with the possibility of eventually completely taking over -- if this happens, the total electrostatic free energy changes sign. We demonstrate this effect on a model of a bacteriophage and summarize our results in Fig.~\ref{fig:pd1}. 

\begin{figure}[h]
\centering
  \includegraphics[width=8cm]{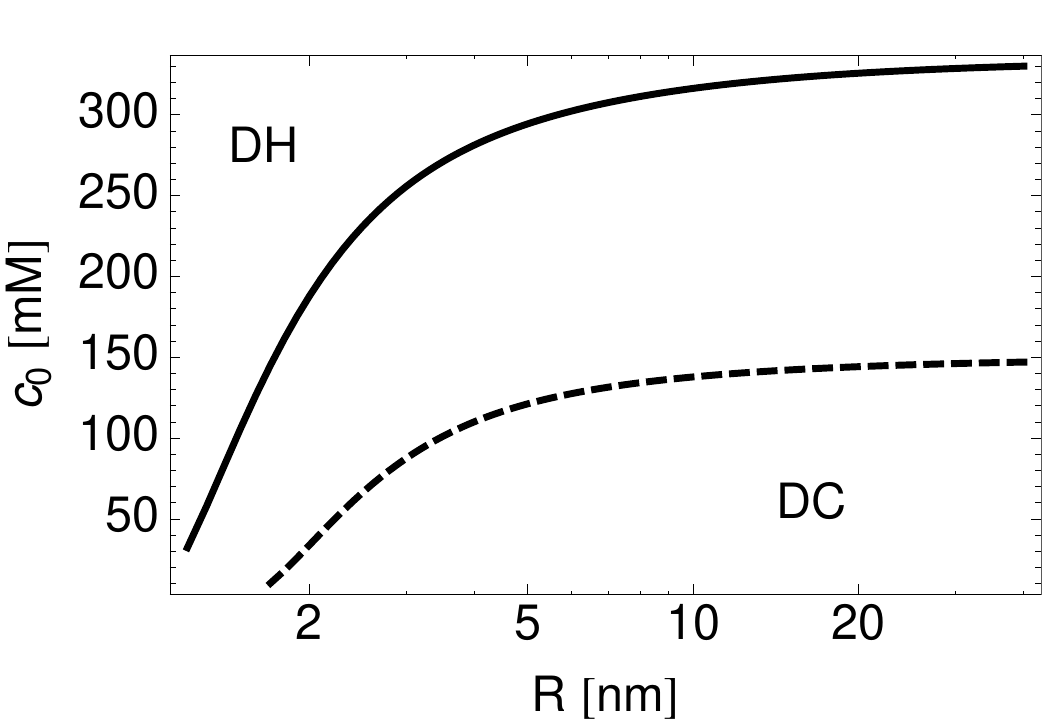}
\caption{The lines separating the positive (above) and negative (below) values of the total free energy of a model bacteriophage with $\rho=-0.7$ $e_0/$nm$^3$, obtained in the dressed counterion approximation. Above the lines the DH term dominates, whereas below the lines the DC term takes over. The bacteriophage is assumed to be in a bathing solution that contains 1 mM of $v$-valent salt, in addition to $c_0$ of mono-valent salt. Full line for $v=4$ and dashed for $v=2$.}
  \label{fig:pd1}
\end{figure}

The boundary line shows where the change of the sign of the free energy happens, i.e. the region below the two curves is where the DH approximation that neglects the non-mean-field effects 
of $v$-valent counterions may fail. At higher monovalent salt concentrations or smaller bacteriophage radii the DC part is much smaller than the leading DH term, meaning that 
the behavior of the total free energy is almost completely determined by it. Below the boundary line, the DC term becomes the dominant one, with the total free energy increasing very rapidly 
with lower salt concentrations or larger radii. Increasing the bulk charge in general means increasing the regime where the DC part dominates. One should, however, exercise some care in 
interpreting these results. The dressed counterion approach is tailored for description of multivalent counterions with $v \gg 1$ and vanishing concentration, so that it may overestimate the effects pertaining 
to 2-valent or even 4-valent counterions. Yet, the model results that we presented do indicate that the strong electrostatic interactions engendered by polyvalent counterions may be important for viruses, especially in the conditions of poor screening by the bathing monovalent salt.

The osmotic pressure of the virion that can be deduced from the above discussion of the free energy behavior can become less repulsive or indeed shows an emerging 
attractive, i.e. negative, component as a function of the volume density of the encapsidated DNA charge, $\rho$. This component has the same origin as the 
attraction observed between two charged macromolecules mediated by polyvalent counterions \cite{naji-rev} and is due to counterion correlations in the strong electrostatic 
field of capsid charges. It thus acts to stabilize the capsid and diminish the osmotic pressure acting within. 

\section{Relevance of physical insight for architecture of real viruses and their "lifecycle"}
\label{sec:relevance}

We have seen that the balance of charges and the electrostatic energy is important for the virus stability, its assembly and function. This physical 
fact restricts the space of virus variability, i.e. the types of proteins encoded by its genome. Of course, 
not all proteins encoded by the genome will be good candidates for a capsid - some of them will not be able to form a capsid 
for sterical reasons. Yet, some of them may easily form empty capsids in the conditions of thermodynamical equilibrium, but 
due to inadequate total charge, or its spatial distribution, cannot encapsidate the RNA molecule. This depends on the amino acid 
content of the virus protein, and the spatial distribution of charge-carrying amino acids. Thus, physical reasons encoded 
in the thermodynamics and free energy of the assembled virus importantly reduce the landscape of possible virus mutants. This 
is especially important in self-assembling ssRNA viruses. In this case  the total charge on the 
ssRNA and in the capsid need to be in a well defined relation \cite{sib_rudi_ssRNA,angelescu_bruinsma,muthukumar} in order to prevent the 
assembly of empty capsids in thermodynamical equilibrium. The stabilizing effects of the nucleic acid in this case transpires also through the negative values of the overall osmotic pressure dur to the polyelectrolyte bridging effects of ssRNA. The length of ssRNA may thus act as a regulator of the capsid size, gathering 
sufficiently large number of proteins in order to screen its charge, i.e. to bring the total charge of the assembled virus 
within the borders enabling a spontaneous assembly (Brownian dynamics studies of ssRNA (generic polymer) virus assembly which account for the electrostatic 
interactions via model pairwise interactions can be found in Ref. \cite{Elrad_Hagan}). The electrostatic interactions between the ssRNA and the proteins can thus 
be viewed as the reason for the characteristic size of the virus \cite{sib_rudi_ssRNA,size_regulation_roya} \footnote{A smilar effect 
has been experimentally found \cite{dragnea1} and theoretically discussed \cite{sib_roya_rudi} in context of assembly of virus proteins around functionalized (negatively charged) nanoscopic golden cores.}.

The calculations we presented predict that there is a polyelectrolyte depleted region in the center of typical ssRNA viruses. 
One may wonder whether this depleted region can be partially or completely  
eliminated in viruses of smaller radii. Such viruses may indeed be viable, but note that they would need to carry more charge on the capsid proteins. Imagine, for example, the evolution of an ssRNA virus whereby its proteins change and instead of e.g. $T=3$ capsid they form a $T=1$ capsid 
\footnote[1]{Some viruses are {\em dimorphic} when assembled without the genome molecule. To observed the dimorphism a modification of the capsid protein is 
sometimes required. C-termini modified (shortened) capsid proteins of hepatitis B virus have been experimentally observed to form $T=3$ and 
$T=4$ capsids \cite{dimorphicHepB}. The percentage of $T=4$ capsid was found to vary, depending on the amount of shortening of the C-termini 
\cite{dimorphicHepB} (see also Ref. \cite{siber_majdandzic}).}. 
Presuming that the ssRNA is not much shortened in this drastic evolutionary event and that it carries similar charge as before, the proteins would need to carry three 
times more charge  in order to pack the ssRNA with the same efficiency as before (here we neglect 
the change in entropic component of the confined ssRNA molecule). It seems reasonable that it would be more difficult to reconcile the stability of 
the fold of such highly charged proteins with the requirement that the proteins, once folded, interact attractively with their neighbors in 
a formed capsid, i.e. that they expose required amino-acids in contact regions \cite{Ptitsyn}. The electrostatics of individual proteins, 
the proteins in capsid and in contact with ssRNA may thus conspire to impose restrictions on virus size.

The total charge on the DNA, i.e. its total length, seems to be a decisive factor concerning the stability of bacteriophages. 
Mutant phages with genomes longer or shorter (deletion mutants) from the wild type virus were studied experimentally (see 
Ref. \cite{taddei_phages} and references therein). Intriguingly, it was found that the deletion mutants were more resistant 
to changes in the environment, particularly to heat shock. The mutant phages with genomes longer from the wild type variant were, 
on the other hand, found to be highly unstable \cite{taddei_phages}. The consideration of electrostatic forces can explain these experimental findings, as 
the internal pressure in the shorter genome (less charged) mutants will be smaller. This may lead to improved mechanical stability of such 
mutants, but this does not necessarily mean that such mutants have evolutionary advantage. Viruses need to be sufficiently 
stable physically, but they must, in addition, {\em (i)} contain all the information needed for their replication, and {\em (ii)} dissasemble 
efficiently in appropriate conditions. One may speculate that 
the longer genomes may offer an evolutionary advantage as they enable the coding of additional proteins, which may be used for 
different purposes, e.g. as scaffolding proteins \cite{scaffold2,Baker_review} enabling a more precise (and possibly quicker) assembly of viruses in infected cells. On 
the other hand, the need for packing the additional ``information'' may render the virus unstable. The virus evolution is thus 
directly influenced by physical constraints on the molecular and nano-scale levels. This is of course true for evolution of all 
organisms, but it becomes particularly transparent in the case of simple macromolecular assemblies such as viruses. 

It was found experimentally that the size and mass of bacteriophage capsids are highly correlated with the genome size \cite{taddei_phages}. 
This can again be explained from the electrostatic considerations, as the phages with longer genomes need larger capsids in order 
to reduce the electrostatic repulsion in the functional (filled) virus. Interestingly, the same type of correlation does not hold for 
ssRNA viruses \cite{taddei_phages,muthukumar}, as the electrostatic interactions there scale differently with capsid radius, and 
the total interior capsid charge is a decisive factor that determines the ssRNA length \cite{muthukumar,sib_rudi_ssRNA} (see Figs. \ref{fig:fig_ssRNA_density} 
and \ref{fig:fig_tails_energies}). The observed differences in ''architectures'' of ssRNA and bacteriophage viruses are also 
a consequence of their ''lifestyles'', in particular the fact that many ssRNA viruses self-assemble, in contrast to bacteriophages which do not.

\section{Conclusions}
\label{sec:conclusions}

We have shown that the size and architecture of viruses are importantly constrained by the (nonspecific) electrostatic interactions acting between 
the constituents of a virus, i.e. its proteins and its genome molecule. A rough understanding of these {\em physical} constraints can be grasped from simple 
mean-field expression for the electrostatic contribution to virus energy (Eqs. (\ref{eq:DH_empty}) and (\ref{eq:bacteriophage_energy})) 
and pressure (Eqs. (\ref{eq:press_ssRNA}) and (\ref{eq:press_DNA})). A more detailed and quantitative argumentation would also have to take into account that electrostatic interactions are only part of the whole picture and that e.g. ion specific effects together with hydration interactions put finer details on
the conclusions reached above. Unfortunately these effects are much less understood and do not allow for simple quantitative estimates as is the case with electrostatic interactions.

Viruses are information encoding machines and most microbiological/virological studies of viruses concentrate on this aspect of viruses (see 
e.g. Ref. \cite{spanish_flu}). Indeed, the viral genome must encode all the proteins required for the life-cycle of a virus and missing in the cellular environment. The typical length of ssRNA genome is often discussed in 
terms of replication fidelity \cite{Holmes,Manrubia}. The rate of mutation of ssRNA viruses \cite{Drake} is about 10$^{-4}$ to 10$^{-5}$ errors per nucleotide 
per round of copying which means that viruses in the daughter generation typically contain one mutation when compared to the parental 
generation \cite{Holmes,Manrubia}. This replication error rate is often considered as the factor restricting the genome length that can be 
faithfully maintained. We have shown that the genome length is also a {\em physically} important quantity that should be examined in context of the size and charge 
distribution of virus capsid (phenotype). This means that the genome length is constrained additionally by the physical interactions conveyed by the 
genome molecule when packed with proteins it encodes. We find here an interesting interplay of non-physical (non-material) and physical/structural 
aspects of the information encoded in a genome molecule that must be tuned properly in a functional virus.

\section*{Acknowledgments}
\label{sec:acknowledgments}

A. \v{S} acknowledges support from the Croatian Ministry of (Grant No. 035-0352828-2837). A. L. B. acknowledges the support from the Slovene Agency for Research and Development. R. P. acknowledges the support from the Slovene Agency for Research and Development, grant no.  P1-0055 and 894.

\footnotesize{
\bibliographystyle{rsc} %
\bibliography{my_references} %
}

\end{document}